\documentclass[aip,graphicx,amsmath,amssymb,preprint]{revtex4-1}

\usepackage{graphicx}
\usepackage{dcolumn}
\usepackage{bm}

\usepackage{xcolor}
\usepackage[normalem]{ulem}

\newcommand{\addition}[1]{#1}
\newcommand{\deletion}[1]{}

\newcommand{\GiPPA}{G_N^{(0,\mathrm{iPPA})}}
\newcommand{\Grheo}{G_N^{(0,\mathrm{rheo})}}
\newcommand{\Gest}{G_N^{(0,\mathrm{est})}}

\begin{document}


\title{Plateau moduli of Kremer-Grest models for commodity polymer melts}

\author{Carsten Svaneborg}
\email{zqex@sdu.dk}
\affiliation{Department of Physics, Chemistry, and Pharmacy. University of Southern Denmark.}

\author{Ralf Everaers}%
\email{ralf.everaers@ens-lyon.fr}
\affiliation{%
 ENSL, CNRS, Laboratoire de Physique (UMR 5672) and Centre Blaise Pascal de l’École Normale Supérieure de Lyon, F-69342 Lyon, France
}%


\date{\today}

\begin{abstract}
We estimate the plateau moduli of \deletion{highly entangled} end-pinned bead-spring polymer melts with $Z=100$ and $Z=200$
\addition{entanglements per chain} from the time-dependent elastic response to a step strain, which we first extrapolate to infinite time and then interpolate to zero strain.
We present data for systems deformed in the melt state as well as for systems deformed at the primitive path level following the recent iPPA protocol. 
We observe excellent agreement between the plateau moduli obtained via the two deformation protocols and good agreement with the available experimental data for commodity polymer melts using a common mapping on the Kuhn scale.
\end{abstract}

\maketitle


\section{\label{sec:Intro}Introduction}

Statistical theories of polymeric systems~\cite{Flory_53,flory1969statistical,degennes79,DoiEdwards86,KhokhlovGrosberg,RubinsteinColby} predict macroscopic properties based on a description and understanding of the microscopic structure and dynamics. 
Scientific progress in the field relies crucially on compilations of microscopic and macroscopic measurements for well-characterized samples.
A classic example is the work by Fetters {\it et al.}~\cite{fetters94}, who assembled  for a wide range of commodity polymers rheological plateau moduli and chain dimensions inferred from scattering experiments.

\addition{Computer simulations offer, in principle, simultaneous access to microscopic and macroscopic properties as well as full control over the sample composition and preparation.}
However, rheological measurements
remain computationally challenging for systems which are as soft and equilibrate as slowly as long-chain polymer melts and solutions. 
Not only must care be taken in properly equilibrating starting states \cite{kremer1990dynamics,auhl2003equilibration,zhang2014equilibration,zhang2015communication,moreira2015direct,SvaneborgEquilibration2016,SvaneborgEquilibration2022}, but long runs and finite deformations are required to measure elastic responses with an acceptable signal-to-noise ratio~\cite{everaers1995test,everaers1999entanglement}.
To make matters worse, the rapidly increasing relaxation times render studies of long chain melts difficult
which
would otherwise help to reduce finite chain length effects or to better separate time or length scale.
\addition{As a consequence and in spite of decades of computational work, we are not aware of the existence of a set of simulation data that is comparable in scope and character to the experimental data assembled by Fetters {\it et al.}~\cite{fetters94}. 
Atomistic \cite{theodorou1986atomistic,kotelyanskii1996building,doherty1998polymerization,faller2001local,abrams2003combined,milano2005mapping,sun2005systematic,tsolou2005detailed,tzoumanekas2006atomistic,harmandaris2006hierarchical,neyertz2008molecular}
and mildly coarse-grain \cite{padding2002time,faller2002modeling,karimi2008fast,eslami2011coarse,chen2008comparison,maurel2015prediction,abdel2004rheological,abrams2003combined,strauch2009coarse,milano2005mapping,maurel2015prediction}
 models can predict sample densities and compressibilities as well as local chain structures and mobilities for given chemical species, but are too expensive to directly extract rheological properties on time scales much beyond the early non-universal glassy regime~\cite{zaccone2023theory}.
Much more coarse-grain slip-link models \cite{hua1998segment,likhtman2005single,NairSchieber2006,ChappaMorseZippeliusMullerPRL2012,UneyamaMasubuchiJCP2012,sgouros2017slip,TheodorouSlipSpring} allow to explore the complete dynamical range for polymer melts in and beyond the linear regime, under both quiescent conditions and under flow. 
However, in our understanding this phenomenological description of the topological effects for linear or branched chains would fail to predict other genuine topology-induced effects like the self-similar territorial structure of melts of non-concatenated ring polymers~\cite{schroeder2026ring}.
}

Here we present \addition{estimates of plateau moduli} 
for a family of bead-spring model polymer melts~\cite{kremer1990dynamics,faller1999local}.
Over the years, the microscopic structure and topology of these systems have been extremely well characterized~\cite{kremer1990dynamics,PPA,Svaneborg2020Characterization}
\addition{and their rheological properties are increasingly studied by direct simulation~\cite{kremer1990dynamics, kroger1993rheology, kroger2000rheological, kroeger2004simple, hou2010stress, cao2015simulating, hsu2016static, oconnor2018relating, xu2018molecular,oconnor2020topological}.}
We combine several techniques to overcome the difficulties of the triple extrapolation to infinite chain length, infinite time and zero strain with the aim to obtain high quality \addition{rheological} estimates of the entanglement or plateau moduli $G_e=\rho k_BT/N_e$ and $G_N^{(0)}=\frac45 G_e$ \cite{DoiEdwards86,LikhtmanMcLeishMM02}:
\begin{enumerate}
    \item We use multi-scale relaxation methods to build well-equilibrated starting states without having to rely on brute-force equilibration~\cite{SvaneborgEquilibration2016,SvaneborgEquilibration2022}. This allows us to study systems composed of chains measuring $Z=100-200$ entanglement lengths and to essentially eliminate finite chain length effects.
     \item We \deletion{tether}\addition{pin} the chain ends to suppress equilibration modes like contour length fluctuations, reptation and constraint release~\cite{DoiEdwards86,LikhtmanMcLeishMM02,hou2010stress}. As a consequence, stress relaxation in deformed systems is limited to the rubber elastic plateau\cite{DoiEdwards86}.
     \item We subject the samples to uni-axial elongations in the range of $0.7\le\lambda\le1.4$.
    \item We infer the equilibrium normal stresses from well-controlled extrapolations to infinite time of two independent sets of data obtained for samples strained either on the melt or the primitive path level~\cite{Svaneborg2024IPPA}.
   \item Finally, we use the Mooney-Rivlin representation\cite{mooney1940theory,rivlin1948large} of equilibrium stresses at finite deformations to infer the elastic response in the limit of vanishing strain.
\end{enumerate}

\addition{We think of the investigated family of bead-spring models as conveniently coarse-grain, numerically efficient yet ``microscopic'' models in the sense that they preserve the physical behavior on the Kuhn scale (chain connectivity, liquid-like packing as well as the mutual impenetrability of the chain backbones and the resulting local preservation of the microscopic topological state) which lies at the origin of the universal aspects of topological effects in linear, branched and ring polymers\cite{tubiana2024topology}.}
Importantly, the present systems can be systematically mapped~\cite{Everaers2020Mapping} onto the polymer melts studied by Fetters {\it et al.} and cover a range of effective segment densities comparable to the available commodity polymers. 
\addition{We thus hope that the present results will serve a similar purpose as reference data in 
\begin{itemize}
    \item investigations of the link between the rheological properties of entangled polymer liquids and (i) the microscopic structure
    \cite{uchida2008viscoelasticity,Milner2020,hoy2020unified,DietzKrogerHoy2022}
     and (ii) the observables\cite{everaers2012topological} of methods like PPA \cite{PPA,sukumaran2005identifying} or Z1+ \cite{kroger2005shortest,kroger2023z1plus} for the analysis of the microscopic topological state and
    \item the parameterization, development and validation of analytical tube models \cite{DoiEdwards86,mcleish02,dealy2018structure}
    and numerical slip-link models \cite{hua1998segment,likhtman2005single,NairSchieber2006,ChappaMorseZippeliusMullerPRL2012,UneyamaMasubuchiJCP2012,sgouros2017slip,TheodorouSlipSpring} describing the link between the microscopic dynamics on the entanglement scale and the rheological properties in and beyond the plateau regime.
\end{itemize}
}

The paper is structured as follows:
in Sect. \ref{sec:theory} we use Likhtman-McLeish theory to analyze how to estimate the plateau modulus from the elastic response of end-pinned melts strained in the melt state and discuss the behavior of melts strained at the primitive path level. 
In Sect. \ref{sec:methods} we present the Kremer-Grest model and the systems we are studying,
as well as the inverse primitive-path analysis method for straining systems at the primitive path level.
In Sect. \ref{sec:results} we present our results for the normal stress relaxation in systems deformed in the melt state and at the primitive path level along with the extrapolation of our data to infinite time and the interpolation of the results to zero strain.
In Sect. \ref{sec:discussion} we compare our simulation results to experimental results and discuss how the configurational relaxation differs in systems deformed in the melt state and at the primitive path level.  
We finish with our conclusions in Sect. \ref{sec:conclusion}.

\begin{figure*}[h!t]
\includegraphics[width=0.9\textwidth]{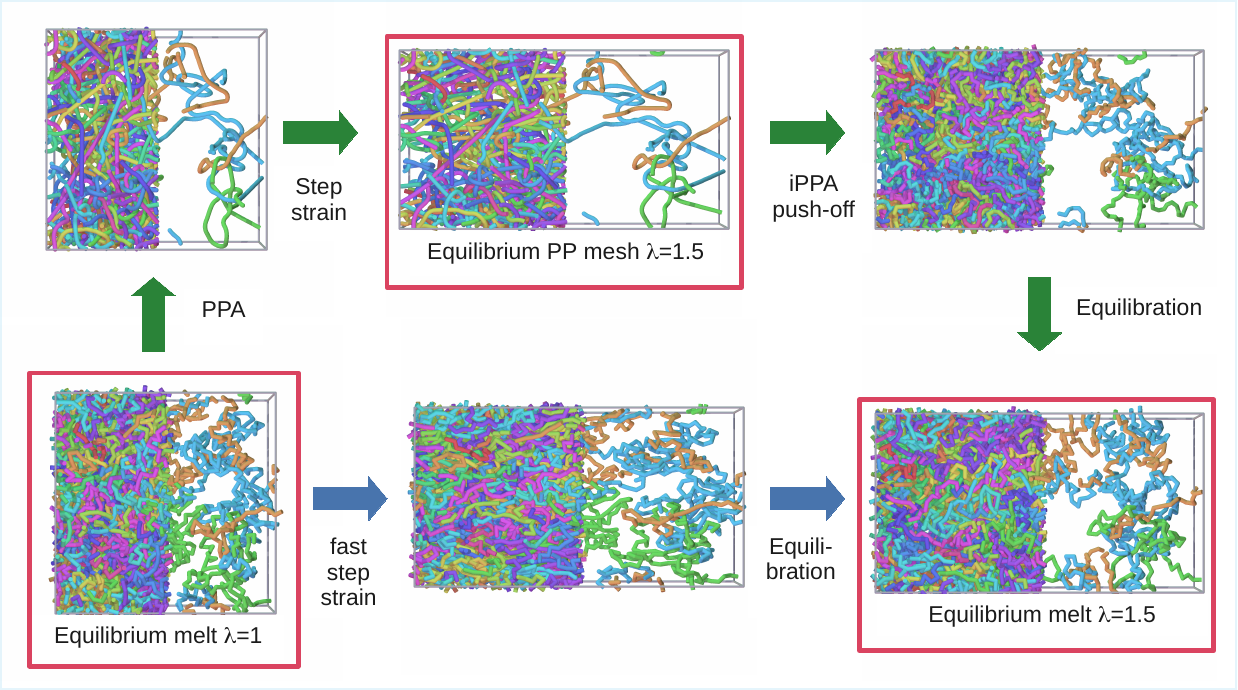}
\caption{\label{fig:iPPAScheme} Illustration of the present methodology. The tiny system visualized here was designed for illustrative purposes only and comprise $M=30$ chains with $Z=3$ entanglements. In each visualization the left side shows the whole melt, and the right side only the same three representative chains.
We use two methods for stress relaxation: \deletion{brute force}\addition{rheological} (blue arrows, bottom row) and primitive-path acceleration (green arrows, top row). The \deletion{brute force}\addition{rheological} approach starts with an unstrained melt state (bottom left), which is rapidly step strained (bottom center), and followed by a very long stress relaxation  at constant strain simulation (bottom right) \addition{corresponding to a rheological step strain experiment}. 
The primitive path accelerated method start\addition{s} by generating the \deletion{equilvalent} \addition{equivalent} primitive path mesh (top left), the mesh is step strained and equilibrated (top center). An inverse PPA push-off is performed whereby excess contour length is reintroduced while switching back to the KG force field (top right). The resulting (unphysical) conformation is equilibrated in a relatively short constant strain simulation.  }
\end{figure*}

\section{\label{sec:theory}Theory}
We infer plateau moduli from the elastic response of long chain model polymer melts with $Z=100$ or $200$ entanglements per chain. We measure this response in two different manners illustrated in Fig.~\ref{fig:iPPAScheme}:
\begin{enumerate}
    \item By subjecting our melts to a rapid uni-axial elongation and subsequently observing the decay of the normal tension in brute-force simulations of the physical relaxation dynamics.
    \item By converting our melts to the corresponding primitive path meshes~\cite{PPA}, straining the primitive path meshes, reverting back to the full Kremer-Grest interactions, and equilibrating the resulting conformation~\cite{Svaneborg2024IPPA}.
\end{enumerate}
Below we discuss the expected time evolution of the normal stresses during the brute-force simulations and the final equilibration phase of the primitive path accelerated relaxation.

\subsection{Shear relaxation modulus for polymer melts with end-pinned chains
\label{sec:TheoryBruteForce}}
The stress relaxation dynamics of a melt with pinned chain ends can be written as a limit of the Likhtman-McLeish theory~\cite{LikhtmanMcLeishMM02} 
\begin{eqnarray}\label{eq:LM}
\frac{G(t)}{G_e} &=& 
\frac{4}{5}+\frac{1}{5Z}\sum_{p=1}^{Z-1}\exp\left(-\frac{p^2t}{\tau_R}\right)
 \nonumber\\
&& + \frac{1}{Z}\sum_{p=Z}^{N_K}\exp\left(-\frac{2p^2t}{\tau_R}\right)
\ .
\end{eqnarray}
where we have eliminated the terms describing stress relaxation via contour length fluctuations, reptation and constraint release.
Here $N_K$ denotes the number of Kuhn segments \deletion{in the polymer}\addition{per chain chain}, $Z=N_K/N_{eK}$ the number of entanglements per chain, $N_{eK}$ the number of Kuhn segments between entanglements, 
\begin{equation}\label{eq:Ge}
G_e = \frac{\rho_K k_BT}{N_{eK}}
\end{equation}
the entanglement modulus, 
$\tau_R = N_K^2 \tau_K$ the Rouse time, 
$\tau_e = N_{eK}^2 \tau_K$ the entanglement time,
$\tau_K$ the microscopic Kuhn time~\cite{Svaneborg2020Characterization,Everaers2020Mapping}, and $\rho_K$ the number density of Kuhn segments. \addition{We expect Eq. (\ref{eq:LM}) to be applicable for describing universal polymer dynamics for times $t \gg \tau_K$.}
Assuming $N_K\gg Z\gg1$ one can replace the discrete sums in Eq. \ref{eq:LM} by integrals.
Using $\mbox{erfc}[x]$ to denote the complementary error function, the resulting approximation 
\begin{eqnarray}\label{eq:LMapprox}
\frac{G(t)}{G_e} &= & 
  \frac{4}{5} \left(1-\frac1Z\right)+ \frac{1}{5}\frac{\sqrt{\pi}}{2}\sqrt{\frac{\tau_e}{t}}
\mbox{erfc}\left(\frac{1}{2}\sqrt{\frac{t}{\tau_R}}\strut\right)
  \nonumber\\
&& + \sqrt{\frac{\pi}{8}}\sqrt{\frac{\tau_e}{t}}\exp\left(-\frac{2t}{\tau_e}\right) 
\end{eqnarray}
is valid for times $t\gg\tau_K$.  

Up to the entanglement time, $\tau_e$, the shear relaxation modulus, $G(t)$, is dominated by the last term in Eqs.~(\ref{eq:LM}) and (\ref{eq:LMapprox}), which describes the Rouse relaxation of short wavelength modes not affected by entanglements. This contribution reduces to zero beyond the entanglement time, $\tau_e$.
At the entanglement time, $\tau_e$, $G(t)$ has decayed to a value of the order of the entanglement modulus, $G_e$, defined in Eq.~(\ref{eq:Ge}). 
The second term describes the relaxation of $1/5$ of the remaining stress  over the Rouse time, $\tau_R$, via the equilibration of longitudinal tension along the tube~\cite{LikhtmanMcLeishMM02}.
For times $\tau_e\ll t \ll \tau_R$ the argument of the complementary error function is small and $\mbox{erfc}[x]\approx 1-2x/\sqrt{\pi}$, so that Eq.~\ref{eq:LMapprox} reduces to
\begin{equation}\label{eq:LMapprox2}
\frac{G(t)}{G_e} 
= \frac{4}{5}\left(1-\frac1Z\right) + \frac{1}{5}\left[\frac{\sqrt{\pi}}{2}\sqrt{\frac{\tau_e}{t}}-\frac{1}{2Z}\right]
\ .
\end{equation}
\addition{Eqs.~(\ref{eq:LMapprox}) and (\ref{eq:LMapprox2}) }
thus suggest two different protocols for estimating 
plateau moduli from the elastic response to a step strain. If the stress relaxation can be followed to completion over times $\tau_R\ll t$, $G_N^{(0)}$ can be read off directly as
\begin{equation}\label{eq:LMPlateauModulus}
\lim_{t\rightarrow\infty}G(t) = G_N^{(0)} = \frac45 \left(1-\frac1Z\right) G_e
\end{equation}
for pinned chain ends with the plateau modulus asymptotically decaying to $4/5$ of the entanglement modulus~\cite{DoiEdwards86}.
%
%

If the stress relaxation can only be followed into the plateau regime, then Eq.~(\ref{eq:LMapprox2}) suggests that data for $G(t)$ from the range $\tau_e\ll t\ll\tau_R$ extrapolate to an apparent plateau modulus, $G_N^{(0,\mathrm{app})}$, when plotted as a function of $1/\sqrt{t}$:
\begin{eqnarray}\label{eq:NeK from Plateau}
G(t) 
&=& G_N^{(0,\mathrm{app})} + \frac{\sqrt{\pi}}{10}\sqrt{\frac{\tau_e}{t}} G_e\\
\label{eq:ApparentPlateauModulus}
G_N^{(0,\mathrm{app})} &=& G_N^{(0)} - \frac1{10Z} G_e
\end{eqnarray}
%
For computational studies this second route has considerable merit, 
since $\tau_R = Z^2 \tau_e$ can exceed the entanglement time by orders of magnitude  for highly entangled chains with $Z\gg1$. 
In particular, the finite chain-length corrections to the asymptotic relation $G_N^{(0)}=G_N^{(0,\mathrm{app})}=\frac45 G_e$ can be made arbitrarily small for larger $Z$ without the need to extend the length of the simulations farther and farther beyond the entanglement time. 

\subsection{Stress relaxation in melts deformed at the primitive path level \label{sec:iPPA Theory}}
As an alternative and potentially faster route towards the equilibrated strained state~\cite{Svaneborg2024IPPA}, we have 
(i) used Primitive Path Analysis (PPA)~\cite{PPA} to convert our end-pinned melts into topologically equivalent meshes of primitive paths,
(ii) deformed the latter at zero temperature (i.e. by continuously minimizing the energy in the course of the deformation), and
(iii) reverted the interactions and temperature back to the original force field and the reference temperature (for details on the implementation see Ref.~\cite{Svaneborg2024IPPA} and Sec.~\ref{sec:PPA force field}). 
Importantly, as in the brute-force simulations of the physical dynamics, the microscopic topological state of the melts is preserved during the entire procedure. 
To sketch an argument comparable in form to the theory in Sec.~\ref{sec:TheoryBruteForce}, below we first discuss the effect of PPA on the chain conformations in mode space.
In a second step, we consider the evolution of the chain conformations in unstrained samples during the  equilibration after switching back to the full KG force field. 
In the final step, we consider the normal stresses during this equilibration for systems that were strained on the primitive path level. 

\subsubsection{From a polymer melt to a mesh of primitive paths \label{sec:theoryPPA}}

PPA contracts the contour of the chains, $L_{pp}\ll L$,  without significantly changing the chain statistics beyond the entanglement scale~\cite{PPA}. 
In particular, the product of contour and Kuhn length for the original chains, $L l_K$, is equal to the product of contour and Kuhn length of the primitive paths, $L_{pp} a_{pp}$, as the mean-square end-to-end distance, $\langle R^2 \rangle$ remains unchanged. 
In mode space, the mean-square excitation $\langle x_p^2\rangle$ of the $p$'th Rouse mode for the chains in the melt~\cite{DoiEdwards86}
\begin{equation}\label{eq:mean square x_p}
    \langle x_p^2 \rangle =  \frac{k_BT}{k_p}
\end{equation}
%
%
due to the equipartition theorem. Here $k_p$ is the spring constant of the $p$'th mode. During PPA it is
reduced to~\cite{everaers1998constrained,mergell2001tube}
%
\begin{equation}\label{eq:mean square X_p} 
    \langle X_p^2 \rangle = \gamma_p\frac{k_BT}{k_p}
\end{equation}
for the randomly quenched~\cite{WarnerEdwards} primitive paths representing their average conformation \deletion{[some
Likhtman reference]}\addition{\cite{likhtman2014tube}}, while the fraction~\cite{everaers1998constrained}
\begin{equation}\label{eq:mean square delta x_p}
    \langle \delta x_p^2 \rangle = \left(1-\gamma_p\right) \frac{k_BT}{k_p} \ .
\end{equation}
of the mode amplitudes removed by PPA corresponds to the annealed fluctuations of the chains within their respective tubes.
The fraction of quenched fluctuations in Eqs.~(\ref{eq:mean square X_p}) and (\ref{eq:mean square delta x_p}) 
is given by~\cite{SvaneborgEquilibration2022}
\begin{equation}\label{eq:gamma_p}
    \gamma_p = \frac{4 Z^2}{\pi^2p^2+4 Z^2}
\end{equation}
and interpolates between $\gamma_p = 0$ for the eliminated short-wavelength modes ($p\gg Z$) and $\gamma_p = 1$ for the long-wavelength modes, which encode the preserved large scale structure ($p\ll Z$). 

\subsubsection{From a mesh of primitive paths to a polymer melt}
After switching back to the original KG force-field the time evolution of $\delta x_p(t) = x_p(t) - X_p$ should be described by a Rouse-like dynamics with~\cite{DoiEdwards86}
\begin{equation}\label{eq:solution Langevin HO}
    \delta x_p(t) = \delta x_p(0) e^{-t/\tau_p} + \int_{0}^{t} e^{-(t-t')/\tau_p} f_{xp}(t')\, dt'
\end{equation}
where the amplitude of the random force $f_{xp}$ is chosen such that Eq.~(\ref{eq:mean square delta x_p}) holds for times $t\gg \tau_p$.
In particular, $\delta x_p(0)=0$ and $x_p(0) = X_p$, since switching back to the original KG force-field only 
leads to a local buckling of the locally smooth primitive paths while the original contour length $L\gg L_{pp}$ is rapidly restored. 
Otherwise we expect the local fluctuations to recover as
\begin{equation}\label{eq:buildup mean square delta x_p}
    \langle \delta x_p^2(t) \rangle = \left(1-\gamma_p\right) \frac{k_BT}{k_p} \left(1-e^{-2t/\tau_p}\right)\ .
\end{equation}
%

\subsubsection{From a strained mesh of primitive paths to a strained polymer melt}
The same restoration of chain fluctuations inside the tube also occurs in samples, where the primitive path mesh has undergone a deformation (which below we take to be a shear deformation for notational simplicity). 
In particular, the shear stress in such a system can be written as the usual mode sum~\cite{DoiEdwards86}
\begin{equation}\label{eq:shear stress from mode sum}
    \sigma_{xy}(t) = \frac{\rho_K}{N_K} \sum_p k_p \langle x_p(t) y_p(t) \rangle
\end{equation}
taking into account that a step shear at $t=0$ transforms the chain coordinates as
\begin{eqnarray*}
    x_p(0^+) &=& x_p(0^-) + \gamma y_p(0^-)\\
    y_p(0^+) &=& y_p(0^-)
\end{eqnarray*}
and that for $t>0$ the $x$- and $y$-coordinates evolve according to equations analogous to Eq.~(\ref{eq:solution Langevin HO}) with uncorrelated noise, $\langle f_{xp}(t) f_{yp}(t) \rangle =0$. 
In particular, $\langle x_p(t) y_p(t) \rangle = \addition{\gamma} \langle y_p^2(0^+) \rangle e^{-2t/\tau_p}$.
Within the standard Rouse model ($\gamma_p=0$ \addition{for all $p$}), Eq.~(\ref{eq:shear stress from mode sum}) reduces \deletion{with} \addition{to} ~\cite{DoiEdwards86}
\begin{eqnarray*}
\label{eq:Rouse G(t)}
   G(t) = \frac{\sigma_{xy}(t)}\gamma  &=&  \frac{\rho_K k_BT}{N_K} \sum_p e^{-2t/\tau_p}
\end{eqnarray*}
\addition{which matches} \deletion{to} the third term in Eq.~(\ref{eq:LM}). 
In the course of the primitive path accelerated relaxation procedure from Ref.~\cite{Svaneborg2024IPPA}, the shear deformation is only applied to the (quenched) primitive path mesh, so that $\langle x_p(t) y_p(t) \rangle = \addition{\gamma}\langle Y_p^2 \rangle$ and 
\begin{eqnarray}
\label{eq:iPPA shear stress}
   G(t) = \frac{\sigma_{xy}(t)}\gamma  &=&  \frac{\rho_K k_BT}{N_K} \sum_p \gamma_p
    \ .
\end{eqnarray}
It is instructive to compare Eq.~(\ref{eq:iPPA shear stress}) to Eq.~(\ref{eq:LM}).
The latter describes the time-dependent elastic response to a step strain deforming the {\it instantaneous} chain conformations in a melt. In particular, the time-independent equilibrium response, Eq.~(\ref{eq:LMPlateauModulus}), is directly observable only after the perturbed annealed parts of all modes have relaxed over their respective characteristic times.
In contrast, straining the primitive path mesh in the iPPA protocol only affects the quenched {\em mean} excitations of the modes.
In this case, Eq.~(\ref{eq:iPPA shear stress}) suggests that the time-independent equilibrium response should be observable almost immediately during the final equilibration phase.
This might seem surprising given that the chain conformations are no less perturbed than in a strained melt, Eq.~(\ref{eq:buildup mean square delta x_p}). The point is that no shear or normal stress are induced as long as the initial local buckling of the primitive paths and the subsequent rebuilding of the annealed fluctuations inside the tube are {\em isotropic}.

\section{\label{sec:methods}Model and Methods}

\subsection{\label{sec:KG model}The Kremer-Grest model}
We simulate Kremer-Grest (KG) model~\cite{grest1986molecular,kremer1990dynamics} polymer melts with an additional bending potential to control the chain stiffness\cite{faller1999local,faller2000local}. 
The KG polymer model~\citep{grest1986molecular} is a bead-spring model where beads interact via 
a Lennard-Jones potential truncated at $r_c=2^{1/6}\sigma$.  This Weeks-Chandler-Andersen\citep{weeks1971role}
(WCA) potential has the functional form

\begin{equation}
U_{WCA}(r)=
\begin{cases}
4\epsilon\left[\left(\frac{\sigma}{r}\right)^{12}-\left(\frac{\sigma}{r}\right)^{6}+\frac{1}{4}\right] & r<r_c \\
0 & r\geq r_c
\end{cases}\label{eq:KG_WCA}
\end{equation}
Bonds are described by the FENE potential
\begin{equation}
U_{FENE}(r)=-\frac{kR^{2}}{2}\ln\left[1-\frac{r^{2}}{R^{2}}\right].\label{eq:KG_FENE}
\end{equation}
With the standard choices of $\epsilon=k_BT$, $R=1.5\sigma$ and $k=30\epsilon/\sigma^{2}$
the average bond length is given by $l_{b}=0.965\sigma$.
The chain stiffness is controlled by a bending potential 

\begin{equation}
U(\Theta)= \kappa\,\epsilon \left(1-\cos\Theta\right),\label{eq:KG_COSINE}
\end{equation}
for the bond angle $\Theta$, which allows to control the chain stiffness. 
The effect of the bending potential is parameterized by a dimensionless parameter $\kappa$, which controls the Kuhn length $l_K(\kappa)$ of our bead-spring chains~\cite{Svaneborg2020Characterization,SvaneborgEquilibration2022}. 
Here we present data for $\kappa\in[0:2.15]$ covering the values relevant to modelling commodity polymers\cite{Everaers2020Mapping}. 
%
%
%
%
To estimate $N_{eK}(\kappa)$ for our systems we use Eq. (60) of Ref. \cite{Svaneborg2020Characterization}, 
for the bead friction $\zeta_b(\kappa)$ we use Eq. (27) of Ref. \cite{Everaers2020Mapping},
and for the Kuhn length $l_K(\kappa)$ we use Eq. (49)  of Ref. \cite{SvaneborgEquilibration2022} with $W=1$. Together these relations provide an estimate $\tau_e(\kappa)$ as well as an estimate for the number of entanglements per chain $Z$ given in Tables~\ref{tab:systems}.

Starting states for the simulated polymer melts with the standard bead density of $\rho_{b}=0.85\sigma^{-3}$ were generated using a recently developed multiscale equilibration method~\cite{SvaneborgEquilibration2022}. 
For integrating the Langevin dynamics of our systems, we used the Gr\o{}nbech-Jensen/Farago (GJF)
integration algorithm\cite{gronbech2013simple,gronbech2014application} as
implemented in the Large Atomic Molecular Massively Parallel Simulator (LAMMPS).\cite{ PlimptonLAMMPS, PlimptonLAMMPS2}
The friction was set to $\Gamma_{b}=0.5m_{b}\tau^{-1}$ and the dynamics integrated with a time step of $\Delta t=0.01\tau$, where $\tau=\sigma\sqrt{m_{b}/\epsilon}$ is the LJ unit of time.


\begin{figure}[th!]
\includegraphics[width=0.7\columnwidth]{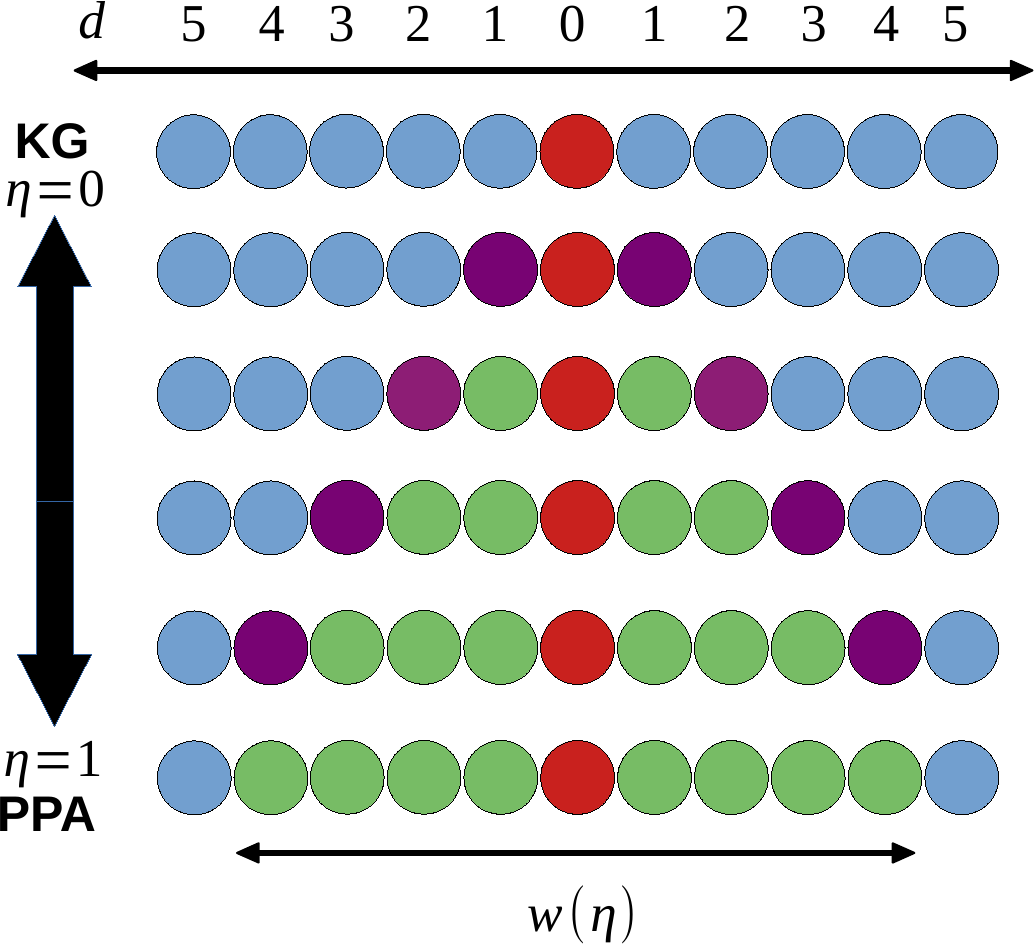}
\caption{\label{fig:ippa}Illustration of the step-wise transformation~\cite{Svaneborg2024IPPA} between the KG and PPA force fields, which allows for a reversible interconversion between topologically equivalent melt states and primitive path meshes. The central red bead has WCA interactions with blue beads (KG, $\eta=0$), no pair interactions with green beads (PPA, $\eta=1$), and force capped interactions with magenta beads to switch WCA interactions on or off (iPPA, $0<\eta<1$). The PPA window in the illustration is $W=4$, in PPA applications $W\propto \sqrt{N_{eK}}$. }
\end{figure}

\subsection{\label{sec:PPA force field}Reversible switching between the KG and the PPA force fields}
The primitive path analysis of the microscopic topological state of entangled polymers~\cite{PPA,sukumaran2005identifying} can be carried out with a variant of the Kremer-Grest force field, where non-bonded intra-chain interactions are disabled for pairs of beads within a PPA window $|i-j|\le W\propto \sqrt{N_{eK}}$.

Switching from the KG to the PPA force field is unproblematic; for example, PPA can be implemented as an energy minimization starting directly from the original melt configuration.
In contrast, using a PPA mesh as \addition{the} starting state for a simulation with the KG force field would not be numerically stable. 
Instead, one of us has developed the iPPA protocol (illustrated in Fig.~\ref{fig:ippa}) for a continuous force field transformation~\cite{Svaneborg2024IPPA} that connects the full KG force field (at $\eta=0$) and the PPA force field (at $\eta=1$) allowing for a reversible force-field transformation or "push-off"~\cite{auhl2003equilibration}, that is numerically stable and preserves topology. 
%
\addition{After the iPPA force-field transformation, the resulting non-equilibrium melt state is simulated with the standard KG force field including the bending potential.} For detailed equations and characterization of the method, we refer to Ref. \cite{Svaneborg2024IPPA}.

\subsection{Deformation and brute-force relaxation of KG melts}

\begin{table}[t]
\caption{\label{tab:systems}
Systems \deletion{deformed in the melt state} studied here. \addition{($\kappa$: chain stiffness and deformation, Prot:
Deformation protocol(s) applied to the system (I: iPPA, R: rheology), $Z$: number of entanglements per chain, $M$: number of chains in the
melt, $N_b$: number of beads per chain, $N_{tot}$: total number of beads, $\tau_e$ and $\tau_R$ estimated
entanglement and Rouse times, respectively.)}
}
\begin{ruledtabular}
\begin{tabular}{c|c|cccc|cc}
$\kappa$ & Prot.  & $Z$     & $M$     &  $N_b$  & $N_{tot}/10^6$  &  $\tau_e/10^3\tau$  & $\tau_R/10^6\tau$ \\
\hline
$-1.0$  & RI  & $100$   & $510$   & $11293$ &  $5.76$   & $23$   & $230$ \\
$-0.50$ & I   & $100$   & $512$   & $10157$ &  $5.20$   & $17$   & $170$ \\
$0.0$   & RI  & $100$   & $517$   & $8408$  &  $4.35$   & $13$   & $130$ \\
$0.0$   & R   & $200$   & $1003$  & $16819$ &  $16.87$  & $13$   & $540$ \\
$0.25$  & I   & $100$   & $522$   & $7338$  &  $3.83$   & $8.0$  & $80$ \\
$0.50$  & I   & $100$   & $528$   & $6237$  &  $3.29$   & $6.0$  & $60$ \\
$0.75$  & I   & $100$   & $538$   & $5155$  &  $2.77$   & $4.4$  & $44$ \\
$1.0$   & RI  & $100$   & $553$   & $4175$  &  $2.31$   & $3.9$  & $39$ \\
$1.0$   & R   & $200$   & $933$   & $8352$  &  $7.79$   & $3.9$  & $160$ \\
$1.25$  & I   & $100$   & $460$   & $3347$  &  $1.54$   & $2.3$  & $23$ \\
$1.50$  & I   & $100$   & $486$   & $2679$  &  $1.30$   & $1.7$  & $17$ \\
$1.70$  & I   & $100$   & $514$   & $2261$  &  $1.16$   & $1.4$  & $14$ \\
$2.0$   & RI  & $100$   & $451$   & $1792$  &  $0.81$   & $0.95$ & $9.5$  \\
$2.0$   & R   & $200$   & $1044$  & $3584$  &  $3.74$   & $0.95$ & $38$  \\
$2.15$  & I   & $100$   & $482$   & $1617$  &  $0.78$   & $0.94$ & $9.4$  \\
\end{tabular}
\end{ruledtabular}
\end{table}


For our first set of results the KG melts (see Table \ref{tab:systems}) were subjected to volume-conserving uni-axial deformations, where they were deformed by a factor of $\lambda=0.9$ or $1.1$ in the parallel direction and by a factor of $1/\sqrt{\lambda}$ in the two perpendicular directions \deletion{during simulations of $100\tau$ }\cite{everaers1995test,everaers1999entanglement}.
\addition{From a theoretical point of view, it would be ideal to implement the strain as an instantaneous deformation. For reasons of numerical stability, the deformation is carried out continuously in quick runs of a length of $100\tau$. The time $t=0$ is set symmetrically to the middle of this deformation period.}
Subsequently, we performed a long stress relaxation simulation at constant deformation, where we have sampled  at every time step all elements of the virial stress tensor $\sigma_{\alpha\beta}(t)$ to measure the time evolution of the normal stress 
\begin{equation}\label{eq:sigma_n(t,lambda)def}
\sigma_n(t,\lambda) = \sigma_{xx}(t,\lambda)-0.5\left(\sigma_{yy}(t,\lambda)+\sigma_{zz}(t,\lambda)\right) \ .
\end{equation}
\addition{We have excluded stress data for $t<50\tau$ from the data analysis, which fall into the second half of the deformation period during which the normal tension is still increasing.}
To deal with the strong fluctuations of the elements of the virial stress tensor, we have averaged $\sigma_{\alpha\beta}(t)$ over time intervals whose width is a multiple of the interval $[\tau_p,\tau_{p+1}]$ between the relaxation time of Rouse modes. With $\tau_p = \tau_K (N_K/p)^2$ these time intervals are equally spaced in representations, where data are plotted against $\sqrt{\tau_{relax}/t}$ for some suitable relaxation time $\tau_{relax}$.

\subsection{Inverse PPA of strained primitive path meshes}
For our second set of results we carried out primitive path analysis (PPA) following the protocol described in Refs.~\cite{PPA,Svaneborg2020Characterization} for $12$ melts with $Z=100$ and $M\approx 500$ for a range of different values of $-1\leq\kappa\leq 2.15$.
Subsequently, we subjected the primitive path meshes to roughly $10$ different uni-axial elongations over the range of $0.7\le\lambda\le1.4$.
Since the contour length redistribution during these deformations occurs via energy minimization, it has a trivial computational cost. 


We then used inverse primitive path analysis (iPPA)~\cite{Svaneborg2024IPPA} as briefly described in Sec.~\ref{sec:PPA force field} to convert the meshes back into topologically equivalent KG melt states.
This was followed by a relaxation simulation of approximately $3 \times10^4\tau$ using the standard KG force field. The times $t$ cited below are counted from the beginning of the last phase.
In particular, we have again sampled the microscopic virial stress tensor during the process with the purpose of extrapolating the data to infinite time corresponding to an equilibrated strained melt.

%
%
%

\subsection{Computational effort and efficiency}

Comparing the computational effort of the rheological simulations and the primitive-path accelerated simulations\addition{,} 
the latter is significantly cheaper at a similar statistical accuracy. 
The PP push off requires $230\tau$ of simulation time, followed by a relaxation simulation which were run
for $10-50$ relaxation times. Thus we used simulations of $10000-50000\tau$ with a median of $32000\tau$
independently of chain stiffness. The overhead of generating the PP mesh of the unstrained melt is amortised
since all deformations start\deletion{s} from the same PP mesh.

For rheological stress relaxation, we performed simulations of $10-10^{3}$ $\tau_e$.
The entanglement time depends strongly on chain stiffness with $\tau_e(\kappa=-1)=23000\tau$, $\tau_e(\kappa=0)=13400\tau$, $\tau_e(\kappa=1)=3870\tau$, and $\tau_e(\kappa=2)=954\tau$.
Hence stiffer melts are much cheaper to simulate compared to flexible melts.
A meaningful comparison is to ask, if we invested the computational effort of a single
PP accelerated simulation, how many $\tau_e$ of rheological stress relaxation would it produce. 
The result is $1.4$, $2.4$, $8.3$, and $34$ $\tau_e$ for $\kappa=-1,0,1,2$, respectively.
Thus our PP accelerated approach is orders of magnitude faster for the flexible systems,
however the computational return is significantly diminished for the stiffer systems.
The present paper comprises approximately $660K$ core hours or $76$ CPU years of
computational effort.

%

\begin{figure}[th!]
\includegraphics[width=\columnwidth]{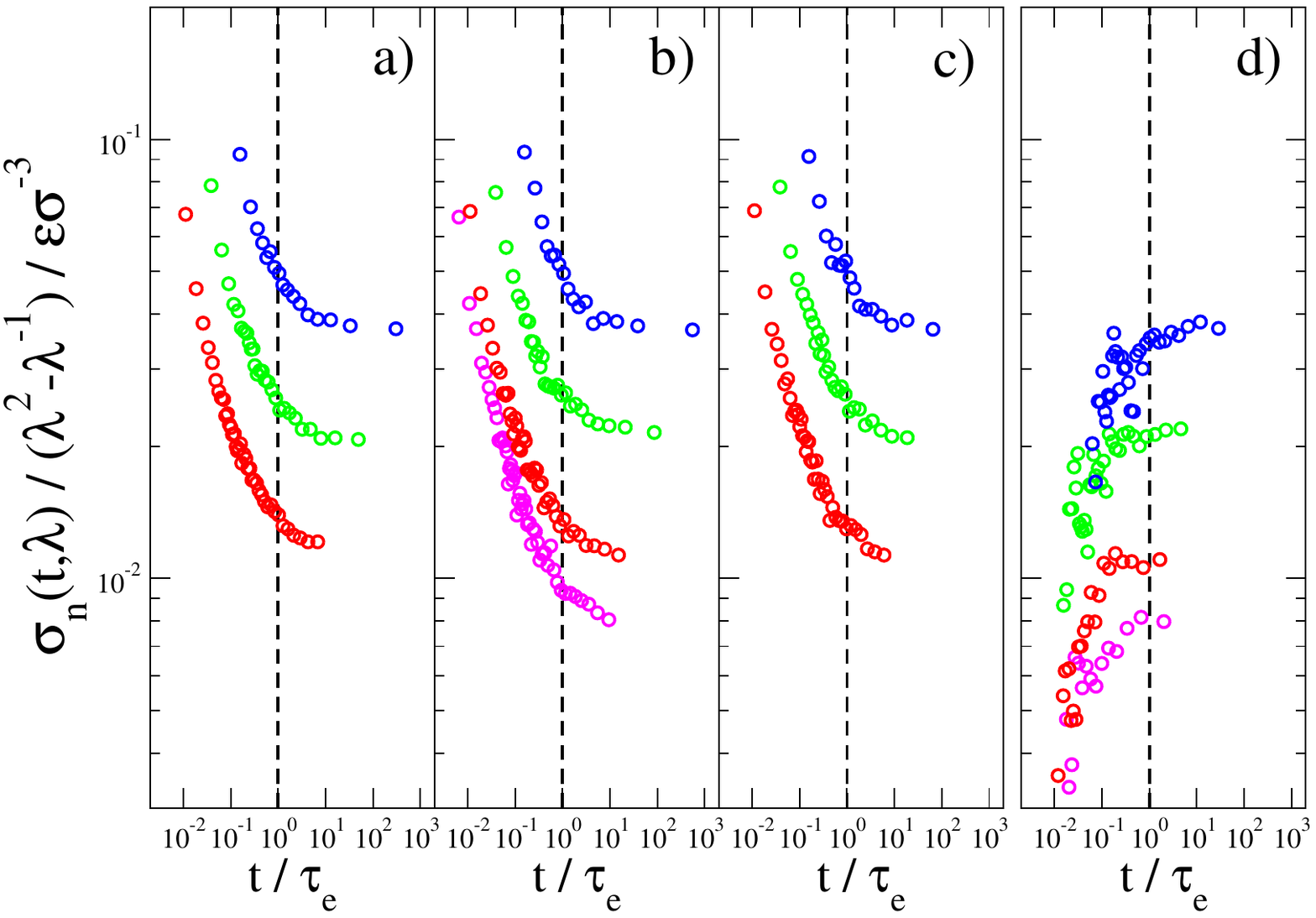}

\caption{\label{fig:stressrelax} 
Relaxation of normal tensions $\sigma_n(t,\lambda)$ in KG melts 
(a-c) after a volume-conserving uni-axial elongation of equilibrated KG melts by a factor of $\lambda=1.1$, (d) after an iPPA pushoff following an equivalent deformation of corresponding  primitive path meshes.
Colors denote the chain stiffness of $\kappa=-1$ (magenta), $\kappa=0$ (red), $\kappa=1$ (green), and $\kappa=2$ (blue).
(a) $Z=200$ and fixed chain ends, (b and d) $Z=100$ and fixed chain ends, (c) $Z=100$ and free chain ends. 
The vertical black dashed line illustrates the entanglement timescale $\tau_e$.
}
\end{figure}

\begin{figure}[th!]
\includegraphics[width=0.95\columnwidth]{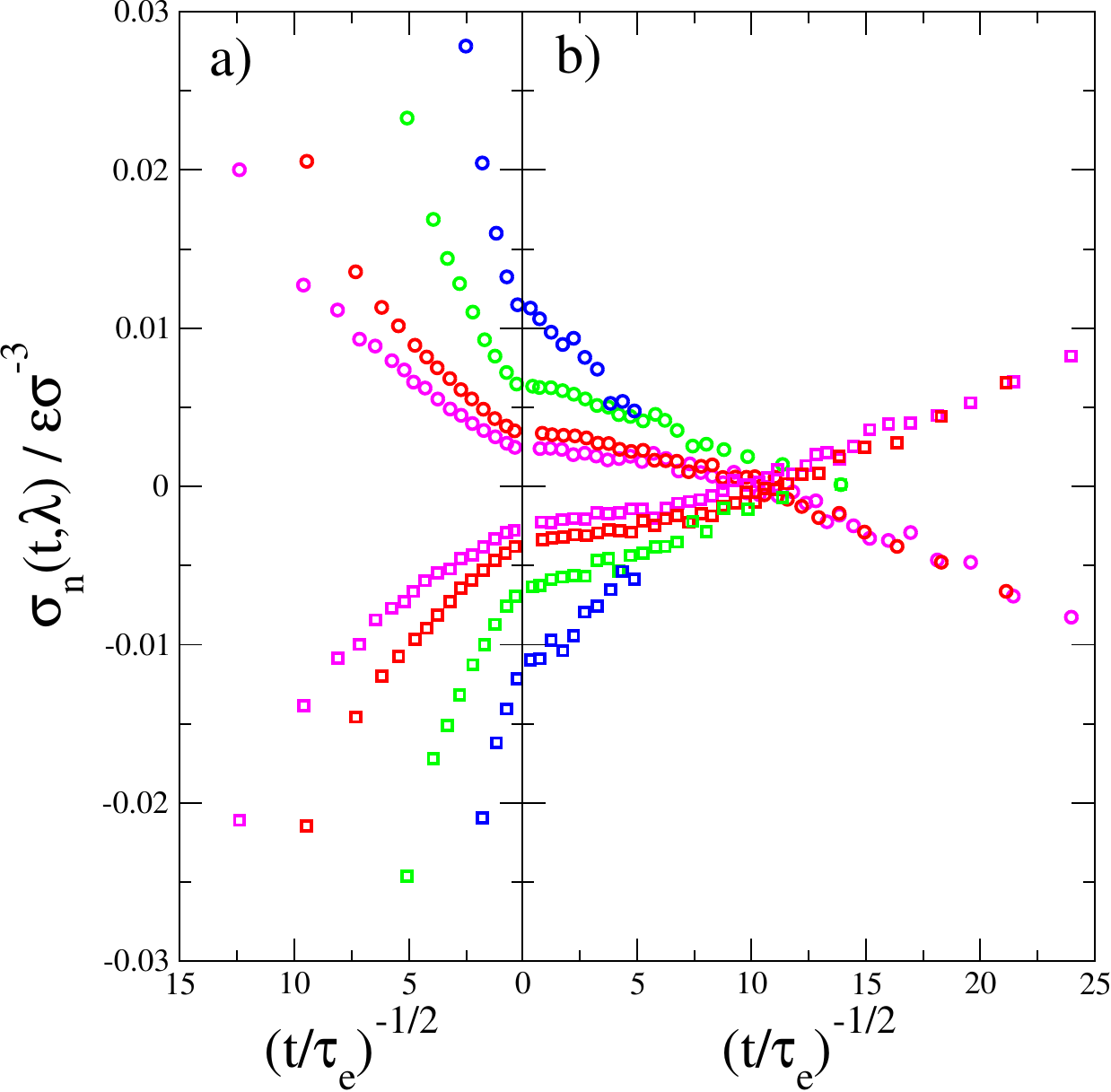}

\caption{\label{fig:stressrelax_versus_1/sqrt(t)}
\addition{
Representation of the same data as in Fig.~\ref{fig:stressrelax} on a linear scale and plotted against $1/\sqrt{t}$ as suggested by Eqs~(\ref{eq:LMapprox}), (\ref{eq:LMapprox2}) and (\ref{eq:NeK from Plateau}): a) systems deformed in the melt state, b) systems deformed at the primitive path level. 
Data are shown for for $t>\tau_K(\kappa)$, symbols denote tension ($\circ$) and compression ($\Box$), and an average has been performed over the data in Fig.~\ref{fig:stressrelax}abc.
Note that we have flipped the time axis in panel a) to have the data extrapolate from two directions to $\lim_{t\rightarrow0}1/\sqrt{t}=0$ in the center. }
}
\end{figure}

\section{\label{sec:results}Results}

\subsection{Temporal evolution of the normal tension}
In Fig.~\ref{fig:stressrelax} we report data for the relaxation of the normal tensions $\sigma_n(t,\lambda)$ in deformed KG melts prepared via our two preparation protocols. 
For our present purposes, the most important observation is that the normal tensions appear to extrapolate to comparable values. Interestingly, they manifestly do so from different directions and possibly on different time scales.

For systems strained in the melt state, the normal tensions exhibit the expected \deletion{monotonous}\addition{monotonic} decay from a high initial value immediately after the step strain. 
In particular, the rubber elastic plateau is reached on the entanglement time, $\tau_e$, while the stress relaxation in the initial Rouse regime is controlled by the Kuhn time, $\tau_K$.
As expected, we observe no discernible differences on the simulated time scales when comparing data for systems with $Z=100$ and $Z=200$  and for systems with fixed and free ends. With a Rouse time of $\tau_R/\tau_e=Z^2=10^4$ ($4\times10^4$) for $Z=100$ ($200$), it would take several orders of magnitude more computational work to reach the Rouse time for these systems.

In contrast, the normal tensions in melts derived via iPPA from strained primitive path meshes appear to reach a plateau after a relaxation time between $10^3\tau$ and $\tau_e$ and typically faster than their homologues deformed in the melt state. Qualitatively, the observed behavior is in agreement with the theoretical arguments from Sec.~\ref{sec:iPPA Theory}. 
Note\addition{, however,} that \deletion{in this case} the normal stress {\it grows} in time instead of decreasing as in the case of systems deformed in the melt state.
\deletion{We tentatively interpret this as a gradual increase of the configurational temperature during the equilibration of the chain statistics, which translates into a reduced effective stiffness of the entropic springs.}

\begin{figure}[th!]

\includegraphics[width=0.95\columnwidth]{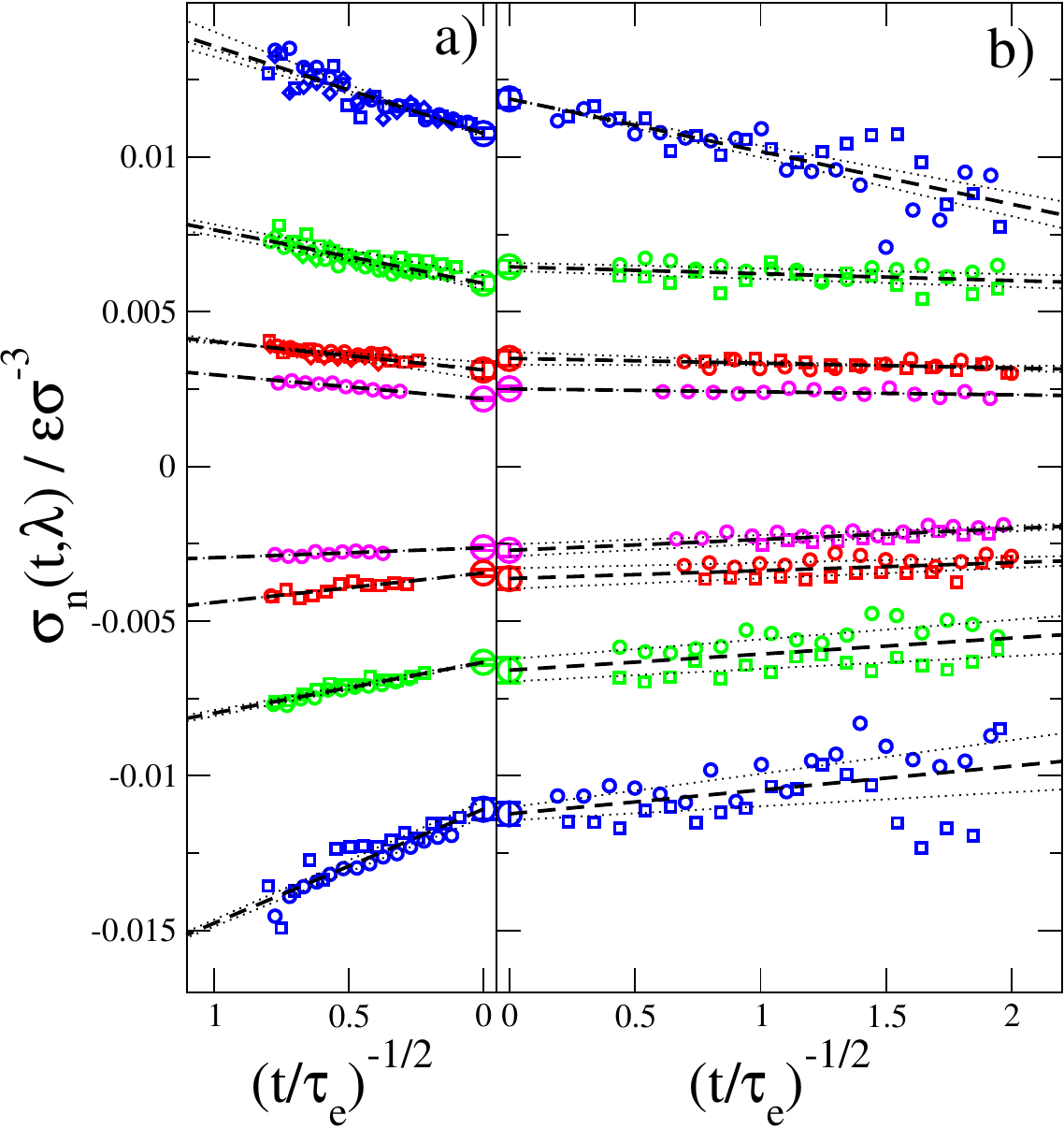}

\caption{\label{fig:stress_extrapolation_to_infinite_time}
Estimation of \addition{$\sigma_n(\kappa,\lambda=1.1)$} \deletion{$G_N^{(0,app)}(\kappa,\lambda=1.1)$(top)}\addition{and $\sigma_n(\kappa,\lambda=0.9)$} \deletion{$G_N^{(0,app)}(\kappa,\lambda=0.9)$ (middle row)} by extrapolation \deletion{of the shear relaxation modulus from}\addition{of the stress data to the limit $\lim_{t\rightarrow0}1/\sqrt{t}=0$ of infinite time for} KG systems with $\kappa=-1$ (magenta\deletion{, *}\addition{ symbols}), $\kappa=0$ (red \deletion{, box}\addition{symbols}), $\kappa=1$ (green\deletion{, diamond}\addition{ symbols}), and $\kappa=2$ (blue \deletion{, triangle}\addition{symbols}).   (For details see the text)
}
\end{figure}

\subsection{Stress extrapolation to the limit of infinite time}
Following the theoretical analysis in Sec.~\ref{sec:theory}, we have \addition{re}plotted the temporal evolution of the normal stresses \deletion{in the plateaus} on a linear scale as a function of $\sqrt{\tau_{e}/t}$ \addition{(Fig.~\ref{fig:stressrelax_versus_1/sqrt(t)})}\deletion{(Fig.~\ref{fig:stress_extrapolation_to_infinite_time})}.
\addition{
For systems deformed in the melt state, the rapidly decreasing early-time normal stresses plausibly extrapolate to zero as predicted by the Rouse model. At the risk of stating the obvious, these data need to be excluded from linear extrapolations to the asymptotic plateau stress. Fits to Eq.~(\ref{eq:NeK from Plateau}) can only rely on (and thus require) data in the plateau regime for times $t\gg\tau_e$.
For systems deformed at the primitive path level, data for times around $\tau_e$ appear sufficient for a robust linear extrapolation, even though the normal tensions do not become completely time-independent as we had hoped originally. 
Remarkably, the linear dependence on $1/\sqrt{t}$ essentially extends all the way down to the earliest relevant times of the order of the Kuhn time $\tau_K$.
Now as the normal tensions increase with time (Fig.~\ref{fig:stressrelax}), they decrease when plotted as a function of $1/\sqrt{t}$.
To our great surprise, for our most flexible systems with $\kappa=-1$ and $\kappa=0$ this decrease persists all the way to normal tensions of {\em opposite} sign but comparable magnitude to what is observed in systems deformed in the melt state.
}
\deletion{
Contrary to our theoretical expectations, the normal tensions in melts deformed at the primitive path level do not become completely time-independent over the simulated time scales. 
Instead they vary over a range comparable to those expected and observed for systems deformed in the melt state.
A post-analysis of our PP meshes revealed that the longitudinal tensions were not fully equilibrated, suggesting that in both cases a similar mechanism might be at the origin of the terminal stress relaxation.
}

\addition{Figure~\ref{fig:stress_extrapolation_to_infinite_time} illustrates how }
we have fitted \deletion{the}\addition{stress} data \deletion{from both data sets to straight lines} \addition{from individual runs around and beyond $\tau_e$} to estimate the normal stresses in the limit of infinite time,
\begin{equation}\label{eq:sigma_n(t,lambda)}
\sigma_n(\lambda) = \lim_{t\rightarrow\infty}\sigma_n(t,\lambda) \ .
\end{equation}
\addition{Corresponding errors were estimated via bootstrapping~\cite{efron1992bootstrap}.}

\begin{figure}[t!]
\includegraphics[width=0.9\columnwidth]{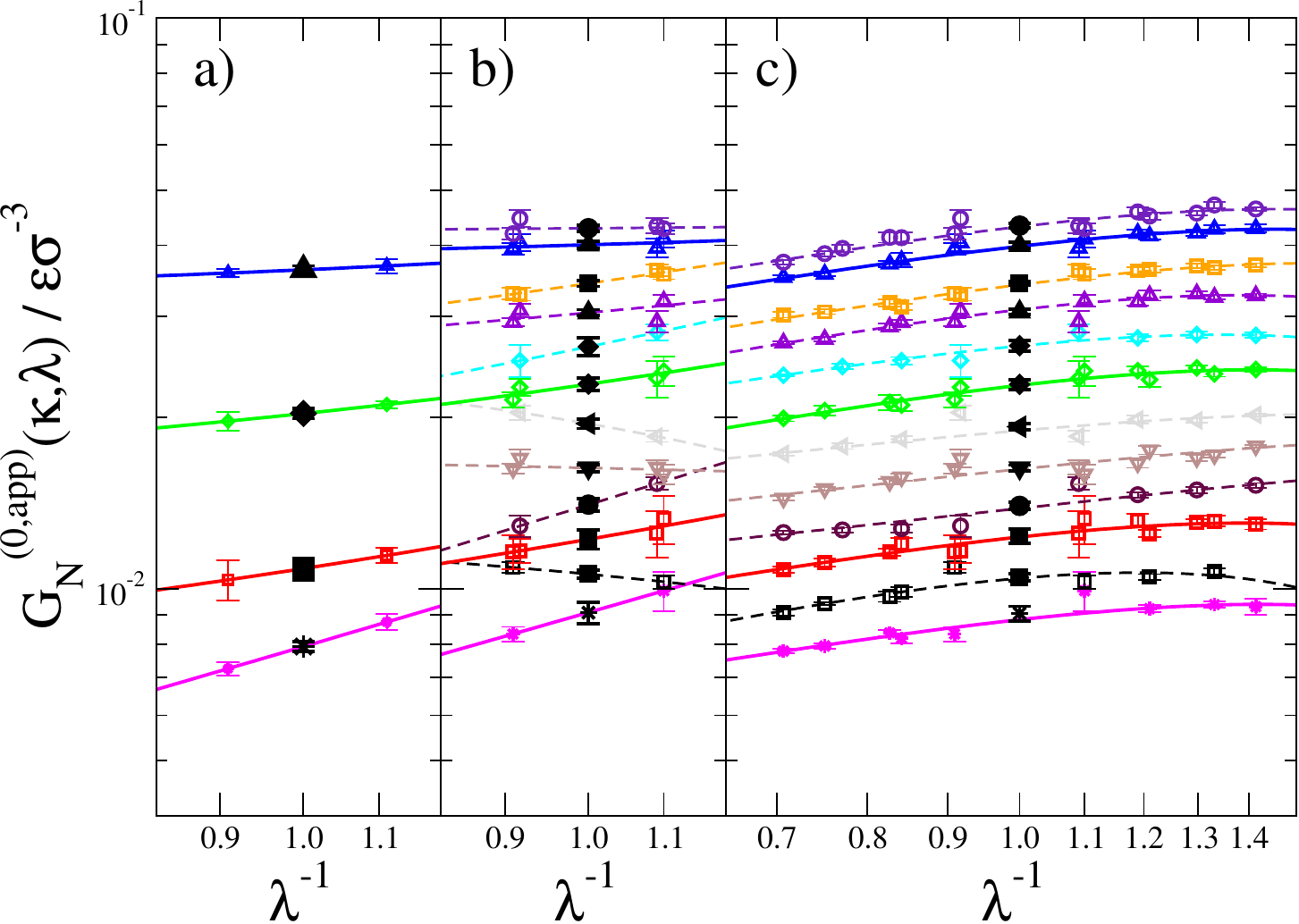}
\caption{\label{fig:mooney} 
\addition{
Estimation of plateau moduli $G_N^{(0,\mathrm{app})}$ (black symbols at $\lambda=1$) 
from extrapolations of \addition{a}symptotic normal stresses to zero strain via the Mooney-Rivlin form Eq.~(\ref{eq:MR}).
(a) systems deformed in the melt state with $\kappa=-1$ (magenta, *), $\kappa=0$ (red, box), $\kappa=1$ (green, diamond), and $\kappa=2$ (blue, triangle).
(b) systems deformed at the PP level with color denoting chain stiffness $\kappa=-1$, $-0.5$, $0$, $0.25$, $0.5$, $1$, $1.25$, $1.5$, $1.7$, $2$, $2.15$ (bottom to top) fitting only data within the same range as in a.
(c) same systems as in b, but where the full deformation range has been fitted.
Systems where we have both deformed the melt state and performed PP acceleration are shown with matching colors and solid lines, additional systems w\addition{h}ere we only have PP accelerated data are illustrated by the hashed lines. 
 \addition{(for details see Sect. \ref{sec:extrapolationtozerostrain})}
}}
\end{figure}

\subsection{Stress extrapolation to zero strain\label{sec:extrapolationtozerostrain}}
%
\addition{Figure~\ref{fig:mooney} shows the strain dependence of the extrapolated normal tensions for the two deformation protocols in a Mooney plot.
Panel a) shows our data for systems deformed in the melt state. 
The two remaining panels show our data for systems deformed at the primitive path level:
in Panel b) restricted to the same $\lambda$-range as the rheological data in Panel a) and in Panel c) over the full range of deformations.
}

\addition{
In this representation, the plateau modulus is given by the limit
\begin{equation}\label{eq:Glimit}
G_N^{(0,\mathrm{app})} = \lim_{\lambda\rightarrow1} \frac{\sigma_n(\lambda)}{\lambda^{2}-\lambda^{-1}} \ .
\end{equation}
}

\addition{To extrapolate the asymptotic normal stresses to zero strain,
we have fitted our data to a power series
\begin{equation}\label{eq:pseudoMR}
G_N^{(0,\mathrm{app})}(\lambda) \equiv \frac{\sigma_n(\lambda)}{\lambda^{2}-\lambda^{-1}} = 2c_1 + \frac{2c_2}\lambda + \frac{2c_3}{\lambda^2}
\end{equation}
from which we can easily read off estimates of the apparent plateau moduli 
\begin{equation}\label{eq:G_MR}
G_N^{(0,\mathrm{app})} = 2c_1 + 2c_2 + 2 c_3 \ .
\end{equation}
Limiting the expansion in Eq.~(\ref{eq:pseudoMR}) to the first order, the coefficients can be identified with those of the standard Mooney-Rivlin form~\cite{mooney1940theory,rivlin1948large}, $c_1=C_1$ and $c_2=C_2$. However, this identification breaks down when including the quadratic term, because the next order in the expansion of invariants introduces three additional terms which contribute not only to $c_3$, but also equal but opposite corrections to $c_1$ and $c_2$. 
}

\addition{
We have carried out fits for large ensembles of synthetic data sets where the values of individual data points are drawn from normal distributions with mean and standard deviations matching our time-extrapolated normal tensions $\sigma_n(\lambda)$.
In Panels a) and b) we show standard Mooney-Rivlin two-parameter fits with $c_3\equiv0$.
Values for $\Grheo$ for systems deformed in the melt state are listed in Table~\ref{tab:rheomoduli} in the supplemental material. 
Corresponding values for $\GiPPA$ for systems deformed at the primitive path level can be found in the second column of Table~\ref{tab:ippamoduli}. 
Corresponding results obtained by including data for all available $\lambda$-values in the Mooney-Rivlin two-parameter fits are listed in the third column of the Table.
As expected, the inclusion of data for larger deformation range
leads to substantially reduced error estimates. However, the small but systematic shift of the estimated plateau moduli exceeds the estimated statistical errors and so
these results cannot be taken as reliable estimates of the limit Eq.~(\ref{eq:Glimit}).
Our best estimates of $\GiPPA$ result from the three-parameter fits of Eq.~(\ref{eq:pseudoMR}) shown in Fig.~\ref{fig:mooney}c) and these are listed in the fourth column
in Tab.~\ref{tab:ippamoduli}.
In this case, we essentially recover the results from the linear fit over the restricted range, but with smaller error estimates.
We refrain from citing numbers for the Mooney coefficients $C_1$ and $C_2$, because linear fits over a restricted deformation range are subject to considerable scatter (Fig.~\ref{fig:mooney}ab) and a correct treatment of non-linear effects ought to be based on a systematic generalization of the Mooney-Rivlin model.
}

\deletion{The nearly stiffness independent slope of the fits suggests $2C_1/G_N\approx 0.75\pm0.04$ and $2C_2/G_N\approx 0.25\pm0.04$. 
The physical origin of the $C_1$ and $C_2$ terms have been discussed at length in the literature in the context of polymer networks see e.g. Ref. \cite{mark1975constants,mark1982use}.
Theories for rubber elasticity make\deletion{s} predictions for the parameters see e.g. Ref. \cite{Gula2020Entanglement} for a discussion. The consensus is that the modulus contribution due to the network structure is entirely described by the $2C_1$ parameter, whereas the entanglement contribution is distributed between the $2 C_1$ and the $2C_2$ parameters. 
There is no consensus between the many theories on what this weighting should be. Nonetheless, the relatively large fraction of the modulus contributed by the $C_1$ parameter is surprising.
}

\addition{
\subsection{Combining estimates of the plateau modulus from the two deformation protocols}
%
}

\begin{table}[t]
\caption{\label{tab:finalmoduli}
\addition{Plateau moduli estimated for the present systems.
($\kappa$: chain stiffness, $l_K$: Kuhn length, $n_K=\rho_K l_K^3$: Kuhn number, and $G_N l_K^3/k_BT$: Kuhn reduced plateau modulus.)
}}
\begin{ruledtabular}
\begin{tabular}{cc|cc|c}
$\kappa$ &  $G_N/\epsilon\sigma^{-3}$  & $l_K/\sigma$  &  $n_K$ & $G_N l_K^3/{k_BT}$ \\
\hline
$-1.00$ & $0.0085\pm0.0006$ & $1.66$ & $2.27$ &  $0.0390\pm0.0027$ \\
$-0.50$ & $0.0100\pm0.0005$ & $1.73$ & $2.45$ &  $0.0512\pm0.0028$ \\
$0.00$ & $0.0116\pm0.0008$ & $1.85$ & $2.80$ &  $0.0730\pm0.0051$ \\
$0.25$ & $0.0133\pm0.0007$ & $1.94$ & $3.08$ &  $0.0965\pm0.0052$ \\
$0.50$ & $0.0155\pm0.0008$ & $2.06$ & $3.47$ &  $0.1344\pm0.0072$ \\
$0.75$ & $0.0183\pm0.0010$ & $2.21$ & $3.99$ &  $0.1962\pm0.0105$ \\
$1.00$ & $0.0215\pm0.0013$ & $2.39$ & $4.69$ &  $0.2949\pm0.0177$ \\
$1.25$ & $0.0253\pm0.0014$ & $2.61$ & $5.60$ &  $0.4518\pm0.0244$ \\
$1.50$ & $0.0290\pm0.0016$ & $2.87$ & $6.76$ &  $0.6872\pm0.0369$ \\
$1.70$ & $0.0326\pm0.0018$ & $3.10$ & $7.90$ &  $0.9735\pm0.0523$ \\
$2.00$ & $0.0381\pm0.0019$ & $3.49$ & $9.99$ &  $1.6208\pm0.0817$ \\
$2.15$ & $0.0411\pm0.0022$ & $3.70$ & $11.23$ &  $2.0837\pm0.1120$ \\
\end{tabular}
\end{ruledtabular}
\end{table}

\addition{
There is no one-to-one correspondence between the data sets we have obtained from the two relaxation protocols. 
Since the rheological protocol of deforming systems in the melt state requires longer runs well into the plateau regime, we have obtained data only for a smaller $\lambda$-range of deformations and a subset of the $\kappa$-values we were able to investigate using the much faster iPPA protocol.
Up to this point in the analysis we have kept the two data sets separate. As a final step, we are left with the task of combining $\GiPPA$ and $\Grheo$
into unified estimates $\Gest$ of the plateau moduli of our various model polymer melts. 
}

\addition{
In principle, $\Gest$ ought to be expressible as a linear combination of $\GiPPA$ and $\Grheo$:
\begin{equation*}
    \Gest = w_\mathrm{iPPA} \GiPPA + w_\mathrm{rheo} \Grheo 
\end{equation*}
with some weights $w_\mathrm{iPPA} + w_\mathrm{rheo} = 1$ and an associated error estimate of
\begin{equation*}
    \delta_{\Gest}^2 = w_\mathrm{iPPA}^2 \delta_{\GiPPA}^2 + w_\mathrm{rheo}^2 \delta^2_{\Grheo} 
\end{equation*}
for $\Gest$.
At this stage we have no reason for putting more faith into one or the other protocol and so one might simply want to choose $w=1/2$. Note, however, that for our data typically $\delta_{\Grheo} \gg \delta_{\GiPPA}$ and so we would end up with {\em less} precise estimates of $\Gest$ whenever we dispose of {\em additional} data from simulations for systems deformed in the melt state.
}
\addition{
Assuming that the two deformation protocols provide two independent and unbiased measurements of $G_N^{(0)}$, this could be avoided by the choice
\begin{eqnarray*}
    w_\mathrm{iPPA} &=& \frac{\delta_{\GiPPA}^{-2}} { \delta_{\GiPPA}^{-2} + \delta_{\Grheo}^{-2} }\\
    w_\mathrm{rheo} &=& \frac{\delta_{\Grheo}^{-2}} { \delta_{\GiPPA}^{-2} + \delta_{\Grheo}^{-2} }
\end{eqnarray*}
which minimizes the estimated error $\delta_{\Gest}$ by suitably increasing the weight of the most precise measurement.
But can we safely assume that our results for $\GiPPA$ and $\Grheo$ are unbiased and subject to independent measurement errors?
}

\begin{figure}
\includegraphics[width=0.95\columnwidth]{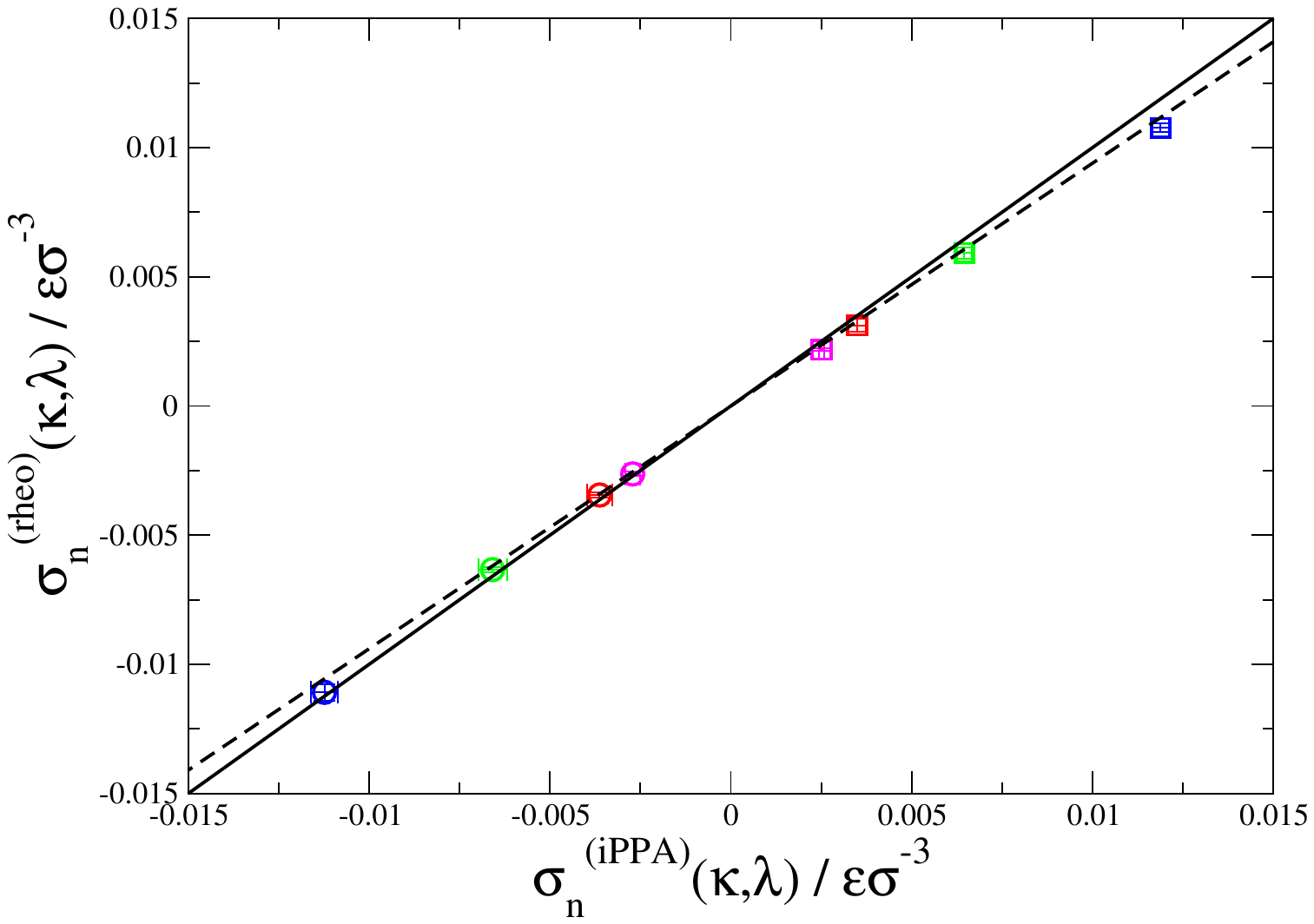}
\caption{\label{fig:stress_extrapolation_correlation}
Correlation of the extrapolation results for $\sigma_N^{(ippa)}(\kappa,\lambda=1.1)$ and $\sigma_N^{(rheo)}(\kappa,\lambda=0.9)$ from Fig.~\ref{fig:stress_extrapolation_to_infinite_time}.
\addition{The lines are $y=\Gamma x$ with $\Gamma=1$ (solid black) and $\Gamma=0.940\pm0.019$ (dashed black).}
}
\end{figure}

\addition{
An inspection of Fig.~\ref{fig:stress_extrapolation_to_infinite_time} suggests that across all four $\kappa$-values available for the comparison the magnitudes of the extrapolated normal tensions are slightly smaller for the rheological protocol compared to the iPPA protocol. 
While these differences are small on the scale of the variation of $G_N^{(0)}$ over the range of the investigated $\kappa$-values, they are larger than our estimates of the statistical errors.
There is thus the possibility that one or both of our data sets suffer from a small systematic bias. (See also Fig. \ref{fig:stress_extrapolation_correlation})
In this case, the prudent choice is to indeed give equal weight to estimates from both deformation protocols, 
\begin{equation}
\label{eq:Gest sum}
    \Gest = \frac12 \GiPPA + \frac12 \Grheo \ ,
\end{equation}
and to account in the error estimate not only for the statistical error, but also for the systematic error associated with the difference $\GiPPA-\Grheo$:
\begin{eqnarray}
\delta_{\Gest}^2 &=& \frac14 \delta_{\GiPPA}^2 + \frac14 \delta^2_{\Grheo} +\nonumber\\
                              && \frac14 \left(\GiPPA-\Grheo \right)^2 
\label{eq:delta for Gest sum}
\end{eqnarray}
}

\begin{figure}
\includegraphics[width=0.95\columnwidth]{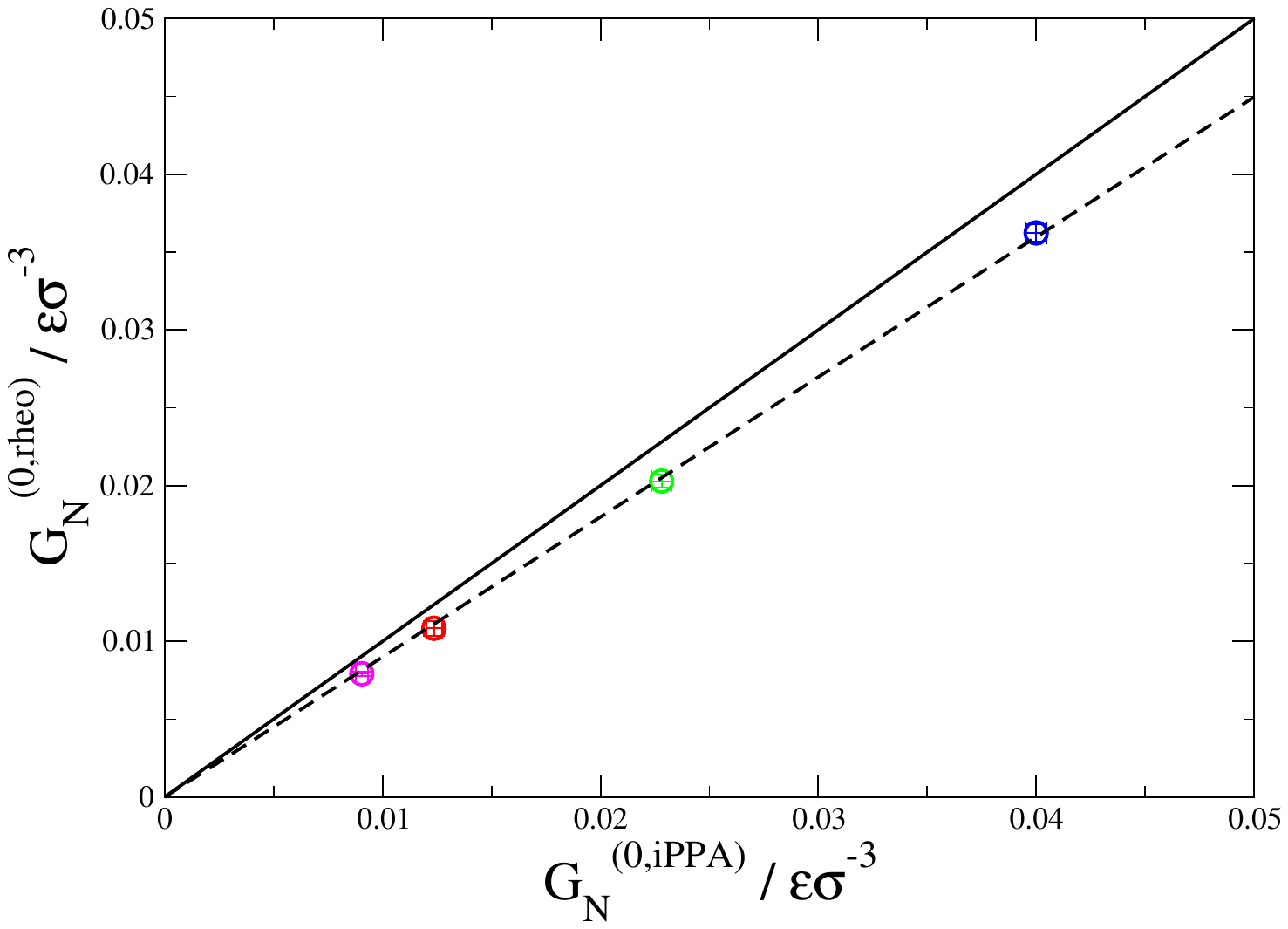}

\caption{\label{fig:moduli_correlation}
Correlation of the interpolation results for $G_N^{(0,app)}(\kappa)$ for the rheological and iPPA protocols  from Fig.~\ref{fig:mooney}.
The lines are $y=\Gamma x$ with $\Gamma=1$ (solid black) and $\Gamma=0.899\pm0.012$ (dashed black).
}
\end{figure}

\addition{
Operationally, Eqs.~(\ref{eq:Gest sum}) and (\ref{eq:delta for Gest sum}) make only sense in cases where we have data from both of the two deformation protocols.
Fig.~\ref{fig:moduli_correlation} suggests that $\Gamma = {\Grheo}/{\GiPPA}= 0.899\pm0.012$ (where the error was estimated by fitting an ensemble of synthetic data
with the same statistics as the simulation data) so that the combined estimate can also be written in the form of a correction to values obtained {\em only} via the iPPA-protocol 
\begin{eqnarray}
\label{eq:Gest from GiPPA}
    \Gest 
      &=& \frac{1+\Gamma}2 \  \GiPPA \ 
\end{eqnarray}
in which case the combined statistical and systematic error can be estimated by
\begin{eqnarray}
\delta_{\Gest}^2 &=& \frac14 \left(1-\Gamma^2\right)\delta_{\GiPPA}^2  + \nonumber\\
&&\frac14  \left(\delta_\Gamma^2 + \left(1-\Gamma\right)^2 \right) {\GiPPA}^2 \ .
\label{eq:delta for Gest from GiPPA}
\end{eqnarray}
The resulting estimates for the plateau moduli of Kremer-Grest melts are listed in Tab.~\ref{tab:finalmoduli}.
Where data for $\kappa=-1,0,1,2$ are calculated using Eqs. (\ref{eq:Gest sum}-\ref{eq:delta for Gest sum}) and
using Eqs. (\ref{eq:Gest from GiPPA}-\ref{eq:delta for Gest from GiPPA}) for the rest. We note that 
the two approaches are in agreement by construction when we have both iPPA and rheological data. We observe
that the relative error in all cases is $5-7$\%, and that it is dominated by the systematic error.
}

\begin{figure}[th!]
\includegraphics[width=0.9\columnwidth]{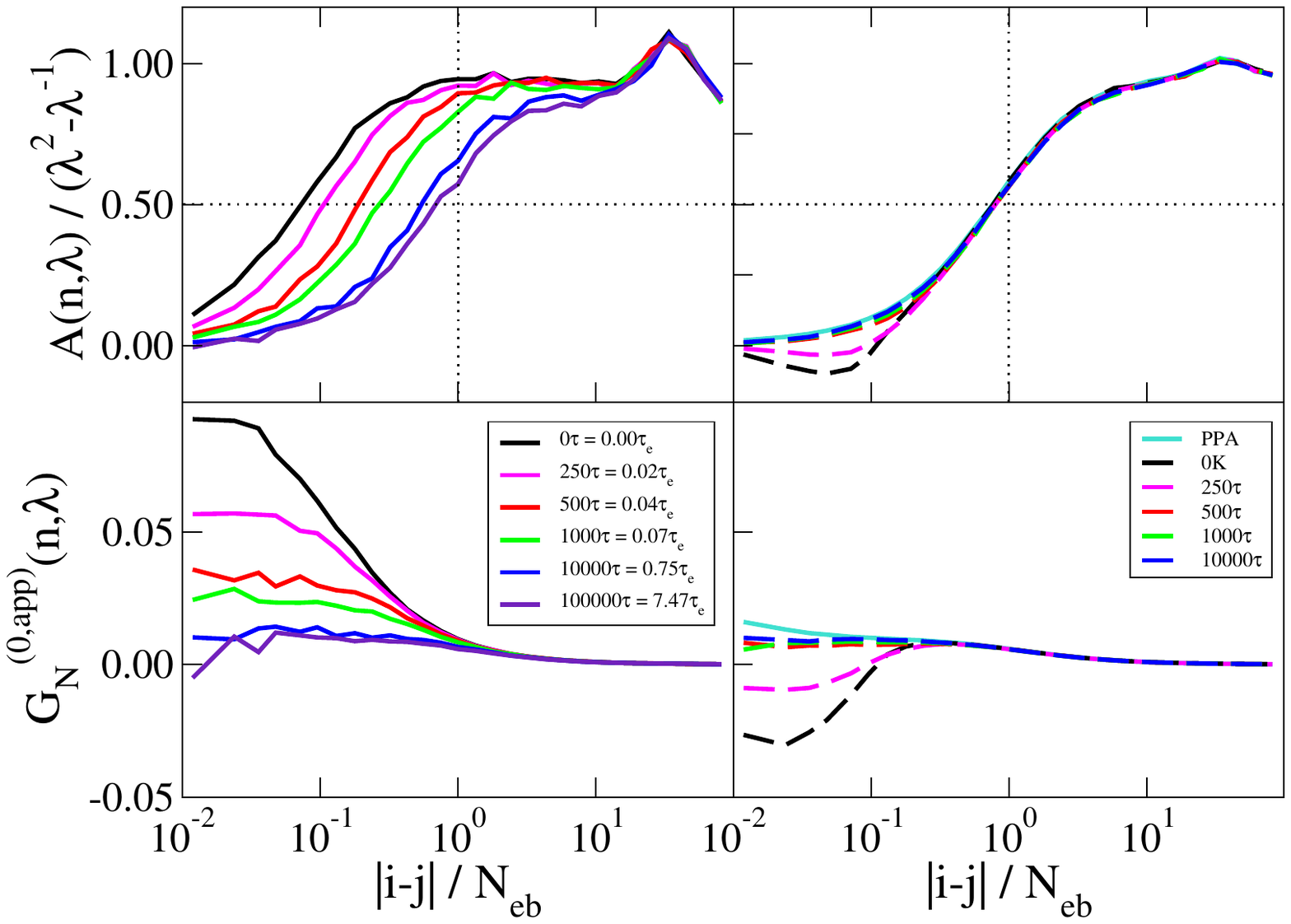}
\caption{\label{fig:anisotropy}
\addition{
Temporal evolution of the chain anisotropy $A(n,\lambda)$  (top row) and modulus $G_N^{(0,app)}(n,\lambda)$ (bottom row)
during relaxation after deformations in the melt state (rheological protocol: left column, $\lambda=1.1$) and on the primitive path level (iPPA protocol: right column, $\lambda=1.5$) for a system with $Z=100$ and $\kappa=0$.
Chemical distances $n=|i-j|$ are shown in units of $N_{eb}=N_{eK}l_K/l_b$, the number of beads per entanglement length.
}
}
\end{figure}

\begin{figure}[th!]
\includegraphics[width=0.95\columnwidth]{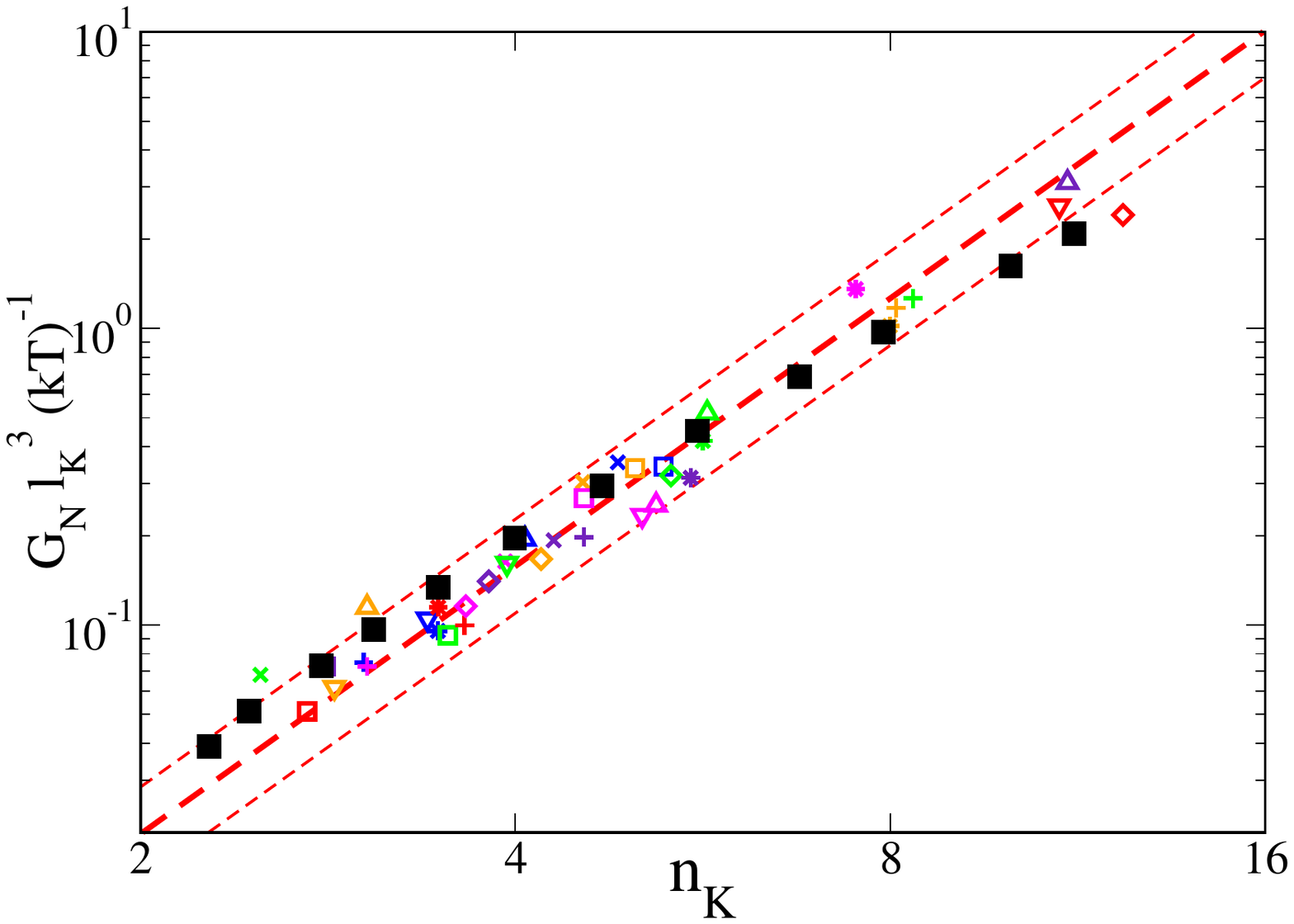}
\caption{\label{fig:experiment}
Comparison between reduced plateau moduli from \deletion{brute force simulations (big black circle), primitive-path accelerated simulations (black filled squares)} \addition{simulations (solid black boxes)} and experimental data for (left to right)
PI-50 (green $\times$),          
PI-7 (red $\Box$),          
PDMS (violet $\Box$),          
PI-20 (orange $\bigtriangledown$),          
PI-34 (blue $+$),          
IPDMS (magenta $+$),          
Icis-PI (orange $\bigtriangleup$),          
cis-PBd (blue $\bigtriangledown$),          
PIB(413) (blue $\ast$),          
cis-PI (red $\ast$),          
a-PP(463) (green $\Box$),          
i-PP (red $+$),          
a-PP(413) (magenta $\diamond$),          
a-PP(348) (violet $\diamond$),          
a-PP (magenta $\times$),          
PIB (green $\bigtriangledown$),          
a-PMMA (blue $\bigtriangleup$),          
i-PS (orange $\diamond$),          
a-PMA (violet $\times$),          
PI-75 (orange $\times$),          
PBd-20 (magenta $\Box$),          
a-PS (violet $+$),          
PBd-98 (blue $\times$),          
PEO (orange $\Box$),          
POM (magenta $\bigtriangledown$),          
a-PHMA (magenta $\bigtriangleup$),          
a-PVA (blue $\Box$),          
SBR (green $\diamond$),          
P6N (violet $\ast$),          
a-P$\alpha$MS (green $\ast$),          
a-PEA (green $\bigtriangleup$),          
PET (magenta $\ast$),          
s-PP (orange $\ast$),          
PE(413) (orange $+$),          
a-POA (green $+$),          
PC (red $\bigtriangledown$),          
PE (violet $\bigtriangleup$),          
PTFE (red $\diamond$),          
For details regarding the polymer abbreviations see Ref. \cite{fetters2007chain}. The thick red dashed line indicates the predictions of the packing argument\cite{lin87,kavassalis87} using $\alpha=18$ and indicating a $20\%$ error by the two thin dashed red lines.
}
\end{figure}

\section{\label{sec:discussion}Discussion}

\subsection{Configurational relaxation and evolution of normal stresses for the two deformation protocols}
The present simulations of long-chain end-pinned polymer melts over the time scale of ${\cal O}(10)$ entanglement times follow the equilibration of our systems over a fraction ${\cal O}(10)/Z^2 = {\cal O}(10^{-3})$ of their longest relaxation time. As a consequence, we are restricted to extrapolating normal tensions to the limit of infinite time, when ideally one would simply measure their equilibrium values.
Extrapolations, even when based on a solid theoretical foundation as the one \addition{exposed} in Sec.~\ref{sec:TheoryBruteForce}, always require a leap of faith and it \deletion{would be}\addition{is} desirable to test their reliability by a comparison of data from independent approaches to the equilibrium state. 
As illustrated in Fig.~\ref{fig:iPPAScheme}\deletion{ and independently of its computational efficiency}, the iPPA protocol~\cite{Svaneborg2024IPPA} provides us\addition{, at least in principle,} with the possibility of doing just that.

\begin{figure}[h!]
\includegraphics[width=0.9\columnwidth]{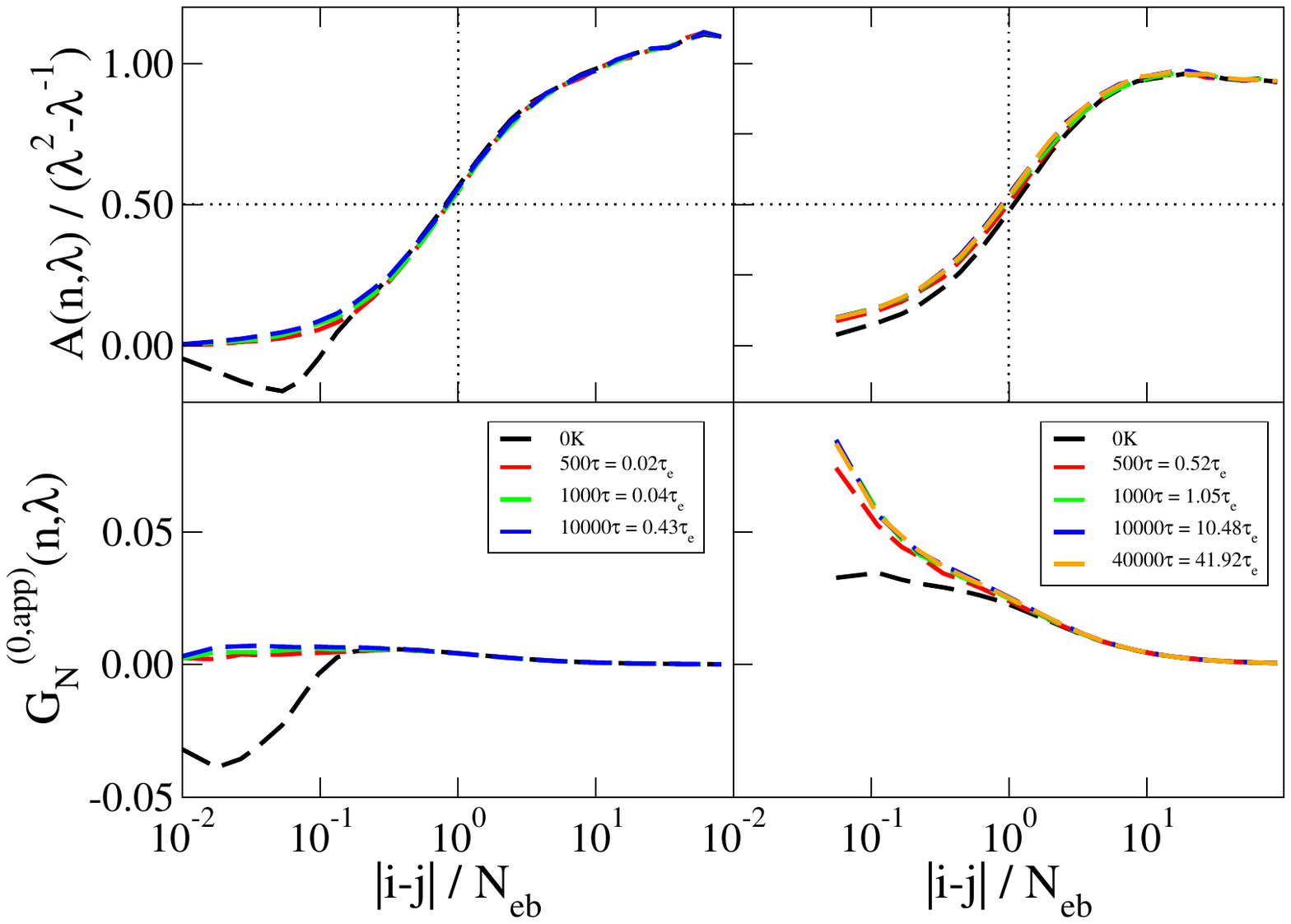}
\caption{\label{fig:anisotropySI}
\addition{
Temporal evolution of the chain anisotropy $A(n,\lambda)$  (top row) and modulus $G_N^{(0,app)}(n,\lambda)$ (bottom row)
during relaxation after deformations on the primitive path level ($\kappa=-1$ left column, $\kappa=2$ right column) for a system with $Z=100$ and $\lambda=1.5$.
Chemical distances $n=|i-j|$ are shown in units of $N_{eb}=N_{eK}l_K/l_b$, the number of beads per entanglement length.
}
}
\end{figure}

\addition{In practice, we are left with several questions.} How different are systems deformed in the melt state and at the primitive path level?
\addition{To which extent do the latter behave as anticipated in Sec.~\ref{sec:iPPA Theory}?
What is the origin of the inversion of the sign of the normal tension reported in Fig.~\ref{fig:stressrelax_versus_1/sqrt(t)} at early times in some of our systems? }
To answer \deletion{this}\addition{these} question\addition{s} it is instructive to \deletion{consider}\addition{compare} the temporal evolution of the configurational statistics on different length scales \addition{for the two deformation protocols. Results for the standard KG model with $\kappa=0$}\deletion{in the two cases} are shown in Fig.~\ref{fig:anisotropy}. 
\deletion{The evolution of chain statistics and chain anisotropy}\addition{Corresponding results} for \addition{iPPA relaxation for} $\kappa=-1$ and $2$ can be found in \addition{Fig. \ref{fig:anisotropySI} of } the Supplementary Information.

To set the stage, consider a solid, where macroscopic strain translates \deletion{instantaneously} all the way down to the scale of its constituent atoms\deletion{ or ions}. As a consequence, their positions change affinely with the macroscopic shape and their distances multiply by the same factors of $\lambda_\alpha$ as the macroscopic dimensions of the sample. In particular, the present uniaxial elongations induce an \deletion{asymmetry}\addition{anisotropy} 
$A(\lambda) = \left(\langle r_{||}^2\rangle - \langle r_\perp^2\rangle\right) / \langle r_\alpha^2\rangle =  \left(\lambda^2-\lambda^{-1} \right)  $
into distances, which were isotropic in the unstrained state, $\langle x^2\rangle = \langle y^2\rangle = \langle z^2\rangle$. 
%
\addition{In a rubber-elastic solid (or in a polymer melt on the time scales considered in the present work) the above scenario applies on large scales. }
\deletion{In a theoretical idealization the above affine deformation describes the situation in a polymer melt or a rubber-elastic solid immediately after an instantaneous step strain. In practice, the length scale dependent chain anisotropy shown in Fig.~\ref{fig:anisotropy}}
\addition{In contrast, deformations} always remain small on the bead \addition{or monomer} scale, which is why systems of this type can sustain large deformations without rupturing the (chemical) bonds along the polymer backbone. 

\addition{
The panels in the top row of Fig.~\ref{fig:anisotropy} show the anisotropy 
\begin{equation}
A(n,\lambda)=\frac{\langle r_{||}^2(n)\rangle - \langle r_\perp^2(n)\rangle}{ \langle r_\alpha^2\addition{(n)}\rangle  }
\end{equation}
of the spatial distances between pairs of monomers $(i,j)$ as a function of their separation $n=|i-j|$ along the chain contour for systems deformed in the melt state (l.h.s.) and at the primitive path level (r.h.s.). The data are normalized to the affine limit, $A(\infty,\lambda) = \lambda^2-\lambda^{-1}$, i.e. the values increase from zero for the quasi-unperturbed bond lengths to one for distant pairs of monomers. 
}
\addition{
Immediately after a step strain in the melt state,
the intrachain distances are affinely deformed almost all the way down to the Kuhn scale. During the subsequent relaxation internal rearrangements slowly increase the crossover scale to the entanglement length.}
\deletion{In particular, the figure shows how the subsequent internal rearrangements at the origin of the decay of the normal tension (Fig.~\ref{fig:stressrelax}) are associated with a gradual reduction of the chain anisotropy on and beyond the entanglement time and length scale. }
\deletion{In Fig.~\ref{fig:anisotropy} we show corresponding data for the temporal evolution of the \addition{reduced} chain anisotropy during the iPPA relaxation following a deformation of the primitive path mesh. }
\addition{Importantly, 
\deletion{the observed behavior is remarkably different as} 
the chain anisotropy in systems deformed in the melt state converges to the chain anisotropy in the deformed primitive path meshes (cyan curve in the top r.h.s. panel of Fig.~\ref{fig:anisotropy}). As anticipated in Sec.~\ref{sec:iPPA Theory}, the anisotropy at and beyond the entanglement scale is not perturbed by the iPPA push-off and remains invariant during the subsequent iPPA equilibration}
\deletion{while}\addition{over time scales where} the overall chain statistics \deletion{continue to}\addition{still} undergo major changes \addition{(see  Fig.~\ref{fig:msid_ratio} in the Supplemental Information)}.
\addition{
However, the iPPA push-off does induce an undesired anisotropy on the Kuhn scale.
While this artificial anisotropy is quickly resolved during the subsequent iPPA equilibration, it has the noteworthy effect to overcompensate rather than reinforce the anisotropy of the primitive path mesh. 
As a potential explanation, consider the local buckling induced by the rapid increase of the contour length during the iPPA push-off (Fig.~\ref{fig:iPPAScheme}), which has a tendency to orient chain segments {\em perpendicular} to the orientation of the original primitive path segment. 
In an isotropic primitive path mesh this has no discernible effect on the chain anisotropy.
But strain orients primitive path segments predominately {\em parallel} to the elongation and so the present form of the iPPA push-off might initially orient chain segments preferentially  in directions {\em perpendicular} to the elongation. 
Note that the effect should be the stronger, the more the length of the chain contour increases during iPPA (and the more the length of the chain contour decreases during PPA). 
For the present systems with $\kappa= \{-1,0,1,2\}$ shown in Fig.~\ref{fig:stressrelax_versus_1/sqrt(t)}, the expected factor decreases with increasing stiffness as $\sqrt{N_{eK}} = \{8.10,6.63,4.10,2.23\}$, giving a first hint, why the artifacts introduced by the iPPA push-off are larger in intrinsically flexible systems. 
(See Fig.~\ref{fig:anisotropySI} of the Supplemental Information)
}

The \deletion{asymmetry}\addition{anisotropy of the chain conformations} lies at the origin of the observed normal tensions.
\deletion{explaining why, as anticipated in Sec.~\ref{sec:iPPA Theory}, the observed normal tensions never exceed the asymptotic values and why they remain essentially constant}
\addition{
Multiplying $A(n,\lambda)$ with $k_BT$ times the density $\rho_b/n$ of entropic springs representing sub-chains of length $n$ yields an estimate of how deformations at and above the contour scale $n$ contribute to the normal tension 
\begin{equation}
\sigma_n(n,\lambda)\approx\frac{\rho_b k_BT}{n} \frac{\langle r_{||}^2(n)\rangle - \langle r_\perp^2(n)\rangle}{ \langle r_\alpha^2(n)\rangle}
\end{equation}
%
with the macroscopic normal tension shown in Fig.~\ref{fig:stressrelax} corresponding to the limit  $\lim_{n\rightarrow0} \sigma_n(n,\lambda)$ \cite{SGE_poly_05}. 
The panels in the bottom row of Fig.~\ref{fig:anisotropy} show $\sigma_n(n,\lambda) = \frac{\rho_b k_BT}{n}A(n,\lambda)$ at different moments in time for the two deformation protocols.
As for $A(n,\lambda)$, we show $\sigma_n(n,\lambda)$ normalized by $(\lambda^2-\lambda^{-1})$ or estimates of how deformations at and above the contour scale $n$ contribute to the shear modulus, Eq.~(\ref{eq:MR}),
\begin{equation}\label{eq:MR}
G_N^{(0,\mathrm{app})}(n,\lambda) \equiv \frac{\sigma_n(n,\lambda)}{\lambda^{2}-\lambda^{-1}} 
\end{equation}
as a convenient way to compare results obtained for different values of $\lambda$.
For systems deformed in the melt state, the local relaxation of the anisotropy corresponds to a slow decrease of the normal tension.
For systems deformed at the primitive path level, the small scale deviations due to the iPPA push-off have a surprisingly large effect.
In particular, the present analysis presents a microscopic interpretation for the surprising {\em inverse sign} and the magnitude of the early-time normal tension and their subsequent {\em increase} towards the asymptotic value reported in Fig.~\ref{fig:stressrelax} and, in particular, Fig.~\ref{fig:stressrelax_versus_1/sqrt(t)}.
}

\deletion{In Fig.~\ref{fig:anisotropy} we show corresponding data for the temporal evolution of the reduced chain anisotropy during the iPPA relaxation following a deformation of the primitive path mesh. 
the observed behavior is remarkably different as the asymmetry converges very quickly after the iPPA push-off  while the overall chain statistics continue to undergo major changes (Fig.~\ref{fig:msid_ratio}).}

\subsection{Comparison to experimentally measured plateau moduli}
Simulation results for Kremer-Grest melts are most conveniently compared to experimental data using Kuhn units~\cite{Everaers2020Mapping, Svaneborg2020Characterization} with the Kuhn length $l_K$ as unit of length, the thermal energy $k_B T$ as unit of energy, and the Kuhn time $\tau_K$ (defined as the time required by a Kuhn segment to diffuse over a distance of $l_K$) as unit of time.
In particular, a  monodisperse polymer sample is characterized by two dimensionless numbers, the chain length $N_K = L/l_K$ expressed in units of Kuhn segments and the Kuhn number $n_K=l_K^3 \rho_K$ measuring density in units of Kuhn segments per Kuhn volume. 

In Fig. \ref{fig:experiment} we report reduced plateau moduli, $G_N^{(0)} l_K^3/k_BT$ as a function of the Kuhn number $n_K=l_K^3 \rho_K$ characterizing the melts.
A first point to note is the excellent agreement between the moduli we have obtained for KG melts by deforming systems either in the melt state ($\bigcirc$) or on the level of the primitive path mesh ($\blacksquare$). 
Secondly, while in good agreement with the experimental data for commodity polymer melts~\cite{fetters1999chain} over the entire relevant range of Kuhn numbers, our results exhibit less scatter. 
For small Kuhn numbers, our data are in good agreement with the prediction of the packing argument\cite{lin87,kavassalis87} giving $N_{eK}=(\alpha/n_K)^2$, where $\alpha$ is the number of chains in the volume of a single entanglement. However, there are systematic deviations towards the upper end of the experimentally relevant range of Kuhn numbers, where the present results confirm our earlier PPA-based predictions~\cite{Svaneborg2020Characterization,Everaers2020Mapping}.
\section{\label{sec:conclusion} Conclusion}

We have taken inspiration from the classic work by Fetters {\it et al.}~\cite{fetters94} to present data for the plateau moduli of a family of microscopically well characterized~\cite{Svaneborg2020Characterization} bead-spring model polymer melts~\cite{kremer1990dynamics,faller1999local} which can be systematically mapped~\cite{Everaers2020Mapping} onto the same commodity polymer melts. 
In a forthcoming paper, we will compare the rheological entanglement lengths~\cite{everaers2012topological} inferred from the present plateau moduli to measures of the entanglement length accessible via PPA~\cite{PPA} or Z1+~\cite{kroger2023z1plus} and test the latters' ability to predict the plateau modulus.

\deletion{Compared to experiment, where}\addition{While} rheological experiments preceded the microscopic sample characterization via neutron scattering by decades\cite{richter2005neutron}, the present results arrive decades after the microscopic structure and topology of Kremer-Grest melts were first investigated \addition{in simulations}~\cite{kremer1990dynamics,PPA}.
\addition{Thus} if the computational results in Fig. \ref{fig:experiment} exhibit less scatter than the experimental data, then \addition{certainly not because rheological properties would be more easily measurable along the present lines. However, we benefit from the fact that all our data were produced following the exact same protocol and} \deletion{the computational approach offers} advantages in the preparation of monodisperse samples as well as in the determination of the microscopic chain dimensions.

The investigation of rheological properties remains computationally challenging, because polymeric systems are soft and equilibrate slowly. 
In particular, brute-force equilibration is much harder to circumvent for problems whose very nature is dynamics than for the equilibration of the static structure~\cite{kremer1990dynamics,auhl2003equilibration,zhang2014equilibration,zhang2015communication,moreira2015direct,SvaneborgEquilibration2016,SvaneborgEquilibration2022}.
The present results suggest that primitive-path accelerated stress relaxation~\cite{Svaneborg2024IPPA} can \deletion{at least} play a useful role for the investigation of \addition{polymer melts and} rubber-elastic systems~\cite{everaers1995test,everaers1999entanglement,svaneborg2004strain,svaneborg2008connectivity,Gula2020Entanglement} under strain.
\addition{
Qualitatively, the relaxation of systems deformed at the primitive path level appears to proceed as we had expected: the asymptotic {\em anisotropy} of the chain conformations is established long before the overall chain conformations are fully equilibrated. 
Quantitatively, we were a little disappointed to find that we still needed to extrapolate the observed normal tensions to infinite times and to detect a $\sim 10\%$ discrepancy between the results obtained from the two deformation protocols.
More work is required to see, if this is just a matter of adjusting the iPPA protocol or if the resolution demands a deeper physical understanding of the primitive path mesh and its response to deformation.
}

\clearpage

\section{\label{sec:acknowledgement}Acknowledgement}
We dedicate this work to Prof. Kurt Kremer on the occasion of his 70th birthday and the 40th anniversary of the publication of Ref.~\cite{grest1986molecular}. It is our pleasure to acknowledge decades of stimulating discussions and interactions with Prof. Kremer and Dr. Grest on the present and related subjects.

The present simulations were performed by using the Large Atomic
Massively Parallel Simulator \cite{PlimptonLAMMPS,PlimptonLAMMPS2}(LAMMPS) Molecular Dynamics software. We acknowledge that part of the results of this research was obtained using the PRACE Research Infrastructure resource Joliot-Curie SKL based in France at GENCI@CEA.
This work was partially supported by DeiC National HPC (g.a. DeiC-SDU-N5-2025141). 

\section{\label{sec:si}Supplementary Material}

Tables of moduli estimated using the rheological and the iPPA protocols
together with their statistical errors obtained via empirical bootstrapping.
In the latter case we have varied both the range of deformation data fitted,
as well as the number of terms in the MR fit.
Correlation plots of stresses and moduli estimated with the rheological
and iPPA protocols. Plots of the evolution of chain statistics and anisotropy
during the relaxation of conformations generated by the iPPA protocol for
several chain stiffness.

\section{\label{sec:data}Data availability}

The melts studied here are a subset of those we have made available from Refs. \cite{SvaneborgEquilibratedKGMeltsZ100, SvaneborgEquilibratedKGMeltsZ200}. Our inverse PPA code has been implemented in the LAMMPS simulator\cite{PlimptonLAMMPS,PlimptonLAMMPS2} and is documented in Ref. \cite{Svaneborg2024IPPA} and can be downloaded from Ref. \cite{ SvaneborgPackageIPPA}

\clearpage
\newpage
\section{\label{sec:sim}Supplementary Material}
\clearpage
\newpage

\begin{table}[t]
\caption{\label{tab:rheomoduli}
Plateau moduli \deletion{and entanglement lengths} extracted from simulations of systems deformed in the melt state.
\addition{($\kappa$: chain stiffness, $G_N^{0,rheo}/\epsilon\sigma^{-3}$: plateau modulus.)}
}
\begin{ruledtabular}
\begin{tabular}{c|cc}
$\kappa$ &  $G_N^{(0,rheo)}/\epsilon\sigma^{-3}$     \\
\hline
$-1.0$   &  $0.00792\pm 0.00016$   \\
$0.0$    &  $0.01084\pm 0.00048$   \\
$1.0$    &  $0.02029\pm 0.00043$   \\
$2.0$    &  $0.03623\pm 0.00058$   \\
\end{tabular}
\end{ruledtabular}
\end{table}

\begin{table}[t]
\caption{\label{tab:ippamoduli}
Plateau moduli extracted from simulations of deformations at the primitive-path level
for $\lambda_{max}^{-1}<\lambda<\lambda_{max}$.
\addition{($\kappa$: chain stiffness, $G_N^{0,rheo}/\epsilon\sigma^{-3}$: plateau modulus.)}
}
\begin{ruledtabular}
\begin{tabular}{c|c|c|c}
$\kappa$ &  $G_N^{(0,iPPA)}/\epsilon\sigma^{-3}$  &  $G_N^{(0,iPPA)}/\epsilon\sigma^{-3}$  &  $G_N^{(0,iPPA)}/\epsilon\sigma^{-3}$    \\
         &  $\lambda_{max}=1.15, c_3\equiv0$ & $\lambda_{max}=1.5, c_3\equiv0$ & $\lambda_{max}=1.5$ \\
\hline
$-1.00$ & $0.00908\pm0.00039$ & $0.00859\pm0.00004$ & $0.00905\pm0.00027$ \\
$-0.50$ & $0.01061\pm0.00018$ & $0.01003\pm0.00005$ & $0.01048\pm0.00014$ \\
$0.00$ & $0.01221\pm0.00047$ & $0.01199\pm0.00006$ & $0.01236\pm0.00036$ \\
$0.25$ & $0.01404\pm0.00034$ & $0.01368\pm0.00005$ & $0.01397\pm0.00021$ \\
$0.50$ & $0.01628\pm0.00023$ & $0.01598\pm0.00005$ & $0.01628\pm0.00019$ \\
$0.75$ & $0.01947\pm0.00037$ & $0.01858\pm0.00007$ & $0.01923\pm0.00023$ \\
$1.00$ & $0.02284\pm0.00059$ & $0.02200\pm0.00012$ & $0.02281\pm0.00045$ \\
$1.25$ & $0.02656\pm0.00097$ & $0.02568\pm0.00010$ & $0.02667\pm0.00056$ \\
$1.50$ & $0.03048\pm0.00037$ & $0.02983\pm0.00009$ & $0.03058\pm0.00033$ \\
$1.70$ & $0.03429\pm0.00040$ & $0.03334\pm0.00011$ & $0.03431\pm0.00029$ \\
$2.00$ & $0.04007\pm0.00063$ & $0.03886\pm0.00016$ & $0.04000\pm0.00049$ \\
$2.15$ & $0.04284\pm0.00055$ & $0.04194\pm0.00016$ & $0.04331\pm0.00044$ \\
\end{tabular}
\end{ruledtabular}
\end{table}

s
\begin{figure}[t]
\includegraphics[width=0.9\columnwidth]{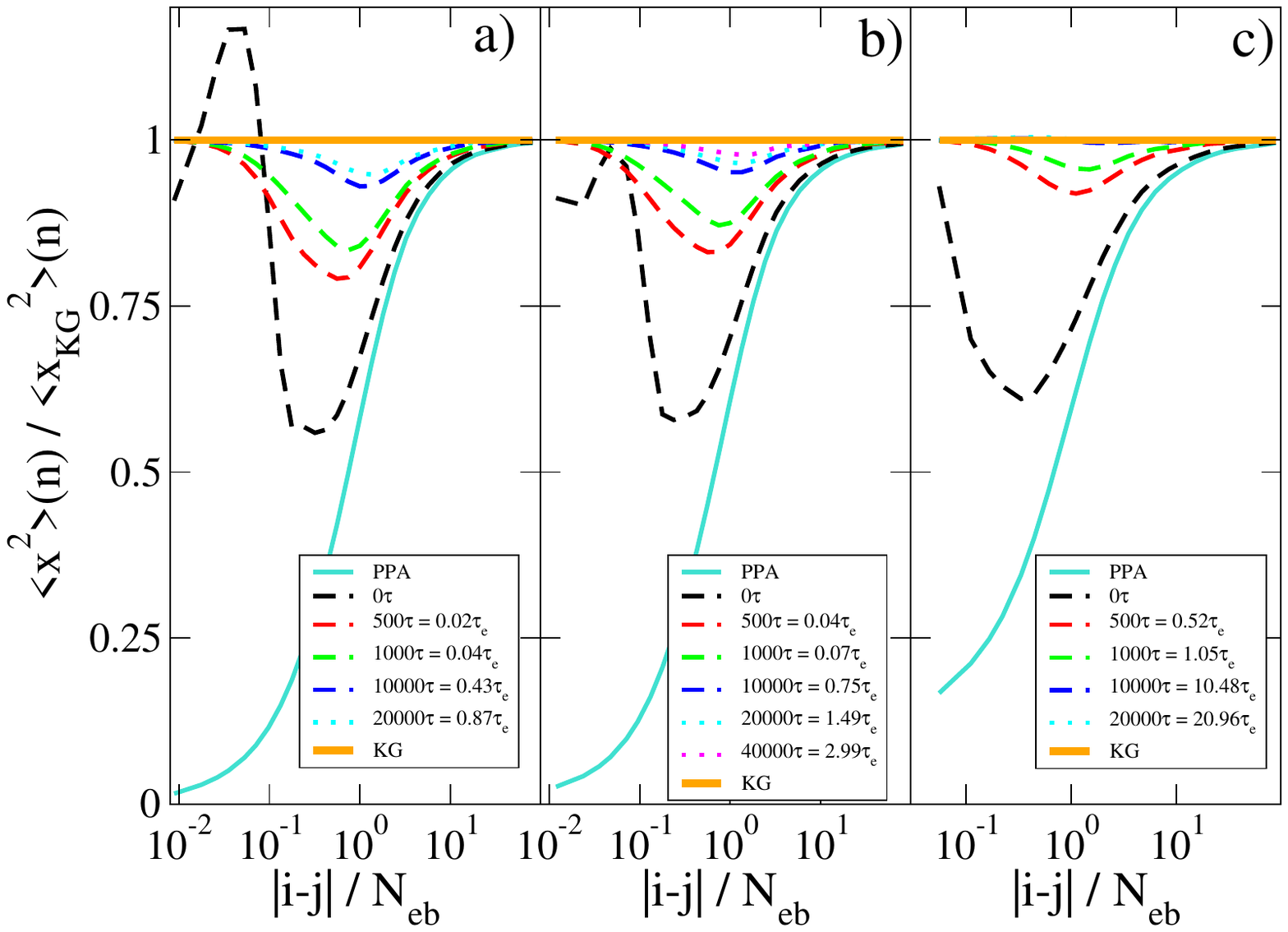}
\caption{\label{fig:msid_ratio}
Temporal evolution of the mean-square spatial distance $\langle x^2\rangle(n)$ 
during the equilibration following the iPPA pushoff for an undeformed melt with $Z=100$ \addition{and $\kappa=-1$ (a), $\kappa=0$ (b), 
and $\kappa=2$ (c)}. 
Data are shown as a function of their chemical distance $n=|i-j|$ along the chain in units of $N_{eb}=N_{eK}l_K/l_b$
the number of beads between entanglements and normalized to $\langle x^2\rangle_{KG}(n)$ for equilibrated Kremer-Grest chains.
Primitive-path mesh (solid \deletion{black}\addition{turquoise} line), conformation just after the iPPA pushoff (black dashed line) and at different times during iPPA equilibration (colored dashed lines \addition{before the entanglement time, and colored dotted lines after the entanglement time}), fully equilibrated KG chains (solid \deletion{red} \addition{orange horizontal} line).
}
\end{figure}

\clearpage
\bibliography{melteq}

\end{document}